\documentclass[prd,aps,twocolumn,a4paper,nofootinbib]{revtex4-1}

\newif\ifusesec
\usesectrue 
   
\usepackage{graphicx,psfrag}
\usepackage{mathrsfs}
\usepackage{amsmath,amsfonts,amssymb}
\usepackage{multirow}
\usepackage{comment}

\DeclareSymbolFontAlphabet{\mathrsfs}{rsfs}
\DeclareMathAlphabet\mathbfcal{OMS}{cmsy}{b}{n}

\newcommand{\be}{\begin{equation}}
\newcommand{\ee}{\end{equation}}
\newcommand{\bea}{\begin{eqnarray}}
\newcommand{\eea}{\end{eqnarray}}
\newcommand{\bel}{\begin{align}}
\newcommand{\eel}{\end{align}}

\def\e{{\rm e}}

\def\GMc2{G M_{\odot} c^{-2}}

\DeclareSymbolFontAlphabet{\mathrsfs}{rsfs}
\DeclareMathAlphabet{\mathcal}{OMS}{cmsy}{m}{n}

\usepackage{color}

\definecolor{cyan}{rgb}{0,0.9,0.9}
\definecolor{orange}{rgb}{0.9,0.5,0}
\definecolor{magenta}{rgb}{1,0,1}
\definecolor{purple}{rgb}{0.8,0.4,0.8}
\definecolor{gray}{rgb}{0.8242,0.8242,0.8242}

\begin{document}

\title{High-order post-Newtonian contributions to the two-body gravitational interaction potential from analytical gravitational self-force calculations}

\author{Donato \surname{Bini}$^1$}
\author{Thibault \surname{Damour}$^2$}

\affiliation{$^1$Istituto per le Applicazioni del Calcolo ``M. Picone'', CNR, I-00185 Rome, Italy}
\affiliation{$^2$Institut des Hautes Etudes Scientifiques, 91440 Bures-sur-Yvette, France}

\date{\today}

\begin{abstract}
We extend the analytical determination of the main radial potential describing (within the effective one-body formalism) the gravitational interaction of two bodies beyond the 4th post-Newtonian approximation recently obtained by us. This extension
is done to linear order in the mass ratio by applying analytical gravitational self-force theory (for a particle in circular orbit
around a Schwarzschild black hole) to Detweiler's gauge-invariant redshift variable.  By using the version of black hole perturbation theory developed by Mano, Suzuki and Takasugi,  we have pushed the analytical determination of the (linear in
mass ratio) radial potential to the 6th post-Newtonian order (passing through 5 and 5.5 post-Newtonian terms). In principle,
our analytical method can be extended to arbitrarily high post-Newtonian orders.
\end{abstract}


\maketitle

\section{Introduction and summary of new results}
\label{sec:intro}

This paper is a follow up of previous work \cite{Bini:2013zaa},  where we analytically determined the radial potential describing (within the effective one-body formalism (EOB) \cite{Buonanno:1998gg,Buonanno:2000ef,Damour:2000we,Damour:2001tu}) the gravitational interaction of two bodies (of masses $(m_1,m_2)$) 
at the 4th post-Newtonian (4PN) approximation.  At this order of approximation,
this determination could be done to all orders in the symmetric mass ratio $\nu :=\mu/M= m_1 m_2 /(m_1+m_2)^2$. (Here $M:=m_1+m_2$ denotes the total mass and $\mu=m_1m_2/(m_1+m_2)$ the reduced mass of the system).
 More precisely,  the 4PN contributions that are nonlinear in $\nu$ were derived from recent
results of Jaranowski and Sch\"afer  \cite{Jaranowski:2013lca}, while the contribution linear in $\nu$ was obtained 
by a combination of techniques. First, we used a recently discovered link \cite{Tiec:2011ab,Tiec:2011dp,Barausse:2011dq} between the $O(\nu)$ piece of the EOB (gauge-invariant) radial potential $A(u;\nu)$ and the $O(\nu)$ piece of Detweiler's gauge-invariant ``redshift'' function $z_1 (\Omega)$ \cite{Detweiler:2008ft}, along a circular orbit of frequency $\Omega$.
 Second, we used a combination of gravitational self-force (GSF) techniques for {\it analytically} computing the $O(\nu)$ piece of $z_1 (\Omega)$, namely: spherical-harmonics-mode-sum regularization \cite{Barack:1999wf,Detweiler:2008ft}, and improved analytic black hole perturbation techniques developed by the Japanese relativity school \cite{Mano:1996vt,Mano:1996mf,Mano:1996gn,Sago:2002fe,Nakano:2003he,Hikida:2004jw,Hikida:2004hs}.

In this paper we give details of our previous work \cite{Bini:2013zaa}, and present the extension of our GSF analysis to
the 6th post-Newtonian (6PN) level (passing through 5 and 5.5 post-Newtonian levels).  Our analysis will give us access
to these high-PN order terms to linear order in $\nu$ only. However, we shall argue that the 5.5 PN contribution is
exactly linear in $\nu$.   
In addition, we suggest that the EOB formalism allows one
to derive some of the coefficients of the higher powers of $\nu$ in the binding energy of a binary system.

We consider a two-body system with masses $m_1$ and $m_2$, moving on a circular
orbit of (areal) radius $r_0$ and orbital frequency $\Omega$, in the
limit where $m_1 \ll m_2$. There are (at least) three ways to express our results. First, it can be expressed in terms of  
the version of Detweiler's gauge-invariant  GSF function \cite{Detweiler:2008ft} that we shall actually compute, namely
\be
 h_{kk}^{R}(u) := h_{\mu\nu}^{R} k^{\mu} k^{\nu} \, .
 \ee
 Here  $h_{\mu\nu}$  denotes the (mass-ratio rescaled) 1GSF metric perturbation of a  Schwarzschild metric  of mass $m_2$:
 $\delta g_{\mu\nu}= g_{\mu\nu} (x;m_1,m_2) - g_{\mu\nu}^{\rm Schwarzschild} (x;m_2) =(m_1/m_2) h_{\mu\nu} + O(m_1^2/m_2^2)$;
   $k^{\mu}$ denotes  the helical Killing vector $k^{\mu} \partial_{\mu} = \partial_t + \Omega \partial_{\varphi}$ of the perturbed metric; and the superscript $R$
 denotes the   {\it regularized}  value of  $h_{\mu\nu}$ on the word line of the small mass $m_1$.  In addition, 
 the argument $u$ in the function  $ h_{kk}^{R}(u) $ denotes  here $G M/c^2 r_0$, 
where $r_0$ is the radius of the considered circular orbit. Equivalently,  one could consider that
 $u=(GM\Omega/c^3)^{2/3}$. 
Indeed, in a first-order  GSF function, such as   $h_{kk}^{R}(u)$, the argument need only
 be defined with background accuracy, i.e. modulo corrections of order $m_1/m_2$.
 In the following,  the independent variable $u$ will be treated as a mathematical argument (varying between
 $0$ and $+ \infty$) whose physical meaning depends on the considered dependent variable.

We shall analytically compute the coefficients entering the 6PN-accurate expansion
(in powers of $u$) of $h_{kk}^{R}(u)$ , i.e. the coefficients $h_n$ in
\begin{eqnarray}
&& h_{kk}^{R}(u)= h_1 u + h_2 u^2 + h_3 u^3 +h_4 u^4 + h_5(\ln u) u^5  \nonumber \\
 & &\quad + h_6(\ln u) u^6 + h_{6.5} u^{13/2} + h_7(\ln u) u^7 + o(u^7) \, .
\end{eqnarray}
Note that a term of order $u^n$ in this expansion corresponds to the $(n-1)$-PN level.  The new terms that we shall
analytically derive below are: (i)  the non-logarithmic part of the 5PN coefficient $h_6(\ln u)$ (its logarithmic part had
been derived in \cite{Blanchet:2010zd}), (ii) the 5.5 PN coefficient  $h_{6.5}$, and (iii) the 6PN coefficient  $h_7(\ln u) $
(which includes a logarithm of $u$).
While we were ready to submit the present manuscript, a preprint by Shah et al. \cite{Shah:2013uya} appeared on the archives,
and reported on an independent analytical derivation of the value of  $h_{6.5}$, as well as on a plausible numerical-analytical
derivation of the logarithmic term in $h_7$. See below for a detailed comparison of our results to the (mainly 
numerical) results of \cite{Shah:2013uya}.

A  second way to phrase our results is in terms of the first-order GSF contribution to the (gauge-invariant) EOB 
 radial interaction potential $A(r;m_1,m_2)= A(u;\nu)$ of a general relativistic two-body system.  
Here, the argument $u$ denotes $u := GM/c^2r$, with  $r$ denoting the 
 (gauge-invariant) EOB radial coordinate. The GSF expansion of  the radial potential $ A(u;\nu)$ reads
\be
\label{eq2a}
A(u;\nu) = 1-2u + \nu a(u) + O(\nu^2)  \, .
\ee
Here the $\nu \to 0$ limiting value $1-2u$  corresponds to the well-known
Schwarzschild potential $A^S (u) = 1-2GM/c^2r =1- 2 u$, while $a(u)$ represents the first-order GSF (1GSF) modification
 of the radial potential  $A(u;\nu) $. 
 
 In this work we shall  use the simple link found in Refs.~\cite{Tiec:2011ab,Tiec:2011dp,Barausse:2011dq}, namely
  \be
  \label{avshkk}
a(u)=  - \frac12 h_{kk}^{R}-\frac{u(1-4u)}{\sqrt{1-3u}}\, ,
 \ee
to deduce from our results on $h_{kk}^R(u)$
 the PN expansion of the 1GSF coefficient $a(u)$ up to the 6PN level, i.e.
 \begin{eqnarray}
 \label{eq2b}
 a(u)&= &a_3 u^3 + a_4 u^4 + a_5(\ln u) u^5 + a_6(\ln u) u^6 \nonumber \\
 &+ &a_{6.5} u^{13/2} + a_7(\ln u) u^7 + o(u^7) \, .
 \end{eqnarray}
 As in the case of the $u$ expansion of   $h_{kk}^{R}(u)$ above,  a term of order $u^n$ in this expansion corresponds to the $(n-1)$-PN level.  The new terms in the expansion (\ref{eq2b}) we shall analytically determine are:
  (i)  the non-logarithmic part of the 5PN coefficient $a_6(\ln u)$ (its logarithmic part had
been derived in  \cite{Damourlogs,Barausse:2011dq}), (ii) the 5.5 PN coefficient  $a_{6.5}$, and (iii) the 6PN coefficient  $a_7(\ln u) $ (which includes a logarithm of $u$).

A third way to express our results is in terms of the (gauge-invariant) function relating the binding energy  $E_B = H^{\rm tot} - M c^2$
of a circular orbit to its orbital frequency  $\Omega$.  Using the convenient dimensionless frequency
parameter  $x := (G M\Omega/c^3)^{2/3}$, it has the form
\begin{eqnarray}
\label{eq10}
E_B (x;\nu) &= &-\frac12 \mu c^2 x (1+e_{\rm 1PN} (\nu) x + e_{\rm 2PN} (\nu) x^2 \nonumber \\
&& + \, e_{\rm 3PN} (\nu) x^3 + e_{\rm 4PN} (\nu , \ln x) x^4 \nonumber \\
&& + \, e_{\rm 5PN} (\nu, \ln x) x^5 + e_{\rm 5.5PN} (\nu)  x^{5.5} \nonumber \\
&& + \, e_{\rm 6PN} (\nu, \ln x) x^6 + o(x^6) )  \, .
\end{eqnarray}
 Below, we shall deduce from our results on the EOB radial potential (using simple EOB-derived results \cite{Buonanno:2000ef,Damour:2009sm}) the analytical expressions of the 1GSF ($\nu$-linear) pieces of   $e_{\rm 5PN} $,
  $e_{\rm 5.5PN} $, and  $e_{\rm 6PN} $.  We shall also indicate several $\nu$-nonlinear terms in the $\nu$-dependent
  coefficients  $e_{\rm nPN}(\nu) $, which arguably follow from the special
  structure of the EOB formalism (within which the $\nu$ dependence of the radial potential $A(u,\nu)$ incorporates
  remarkable cancellations, not present in  $ E_B (x;\nu) $).
  
  The presence of  (conservative) contributions at the $5.5$ PN level (i.e. $O(1/c^{11})$)
   in the various functions mentioned above seems
to conflict with the usual PN lore that conservative (time-even) effects arise at even powers of $1/c$, and that odd powers
of $1/c$ are associated with time-odd radiation-reaction effects .  
As we shall discuss  in detail in Sec. IIIB below,  this unexpected result can be heuristically understood
as stemming  from some results of Blanchet and Damour \cite{Blanchet:1987wq,Blanchet:1992br}
about  hereditary effects (given by an integral over the past behavior of the source) in the {\it inner metric} of the system,
and their relation (via energy balance) to corresponding hereditary effects (``radiative tail'')  in the {\it wave-zone} metric.
We shall find  that our analytical results for the 5.5PN-level (i.e. $O(1/c^{11})$) quantities $h_{6.5}$ and $a_{6.5}$
arise from the inner-metric image of the second-order radiative tail \cite{Blanchet:1997jj}. 
The important conceptual point here is (as already emphasized in  \cite{Blanchet:1987wq}) that the hereditary
tail effects are time dissymetric without being time anti-symmetric. They can thereby generate terms that
are either radiation-reaction-like or conservative (thereby violating the usual PN lore).
[After the submission of the present work, there appeared on the archives a paper by Blanchet et al. \cite{Blanchet:2013txa}
which confirmed our (heuristic) analysis by  a detailed multipolar-post-Minkowskian treatment of the gravitational field outside the source.]

\section{Analytical computation of conservative GSF effects along circular orbits}
\label{sec:two}

\subsection{General strategy for computing  $h_{kk}^R $ }

Detweiler \cite{Detweiler:2008ft} has pointed out the existence of one simple, (conservative) gauge-invariant function, available within first-order $(O(\nu))$ GSF theory, associated with the sequence of {\it circular} orbits of an extreme mass-ratio binary system: $m_1 \ll m_2$. Computing the $O(\nu)$ piece of this redshift function $z_1 (\Omega ; \nu) \equiv 1/u^t_1 (\Omega;\nu)$ is equivalent \cite{Detweiler:2008ft,Blanchet:2009sd,Akcay:2012ea} to computing the {\it regularized} value, along the world line $y_1^{\mu}$ of the small mass $m_1$, of the double contraction of the $O(m_1/m_2)$ metric perturbation $h_{\mu\nu}$ (considered in an {\it asymptotically flat} gauge)
\be
\label{eq18}
g_{\mu\nu} (x;m_1,m_2) = g_{\mu\nu}^{(0)} (x;m_2) + \frac{m_1}{m_2} h_{\mu\nu} (x) + O\left( \left(\frac{m_1}{m_2}\right)^2\right)\, 
\ee
(where $g_{\mu\nu}^{(0)} (x;m_2) \!= \!g_{\mu\nu} (x;m_1 \!= \!0 , m_2)$ is a Schwarzschild metric of mass $m_2$), with the four-velocity $u_1^{\mu} = dy_1^{\mu} / ds_1$ of $m_1$, say
\be
\label{eq18bis}
h_{uu}^R := {\rm Reg}_{x \to y_1} [h_{\mu\nu} (x) u_1^{\mu} u_1^{\nu}] \, .
\ee
Following Refs.~\cite{Barack:1999wf,Hikida:2004jw,Detweiler:2008ft,Blanchet:2009sd} the regularization operation indicated in Eq.~(\ref{eq18bis}) is done by subtracting the (leading-order) singular part in the spherical harmonics expansion of $h_{uu}$. This yields $h_{uu}^R$ as a series indexed by $l=0,1,2,\ldots$
\be
\label{eq19}
h_{uu}^R = \sum_{l=0}^{\infty} (h_{uu}^{(l)} - D_0) \, , 
\ee
where
\be
\label{eq20}
h_{uu}^{(l)} = \sum_{m=-l}^{+l} u^{\mu} u^{\nu} h_{\mu\nu}^{(l,m)} \, , 
\ee
and where the ($l$-independent) subtraction constant $D_0$ is known \cite{Detweiler:2002gi,Detweiler:2008ft} to be
\begin{eqnarray}
\label{eq21}
D_0 &= &2u\sqrt{\frac{1-3u}{1-2u}} F \left( \frac12 , \frac12 , 1 ; \frac{u}{1-2u} \right) \nonumber \\
&= &2u \sqrt{\frac{1-3u}{1-2u}} \left( \frac2\pi K(k) \right) \, .
\end{eqnarray}
Here, $u = Gm_2 / c^2 r_0 = GM/c^2 r_0 + O(\nu)$, $F(a,b,c;z)$ is Gauss's hypergeometric function and $K(k) = \int_0^{\pi/2} d\theta / \sqrt{1-k^2 \sin^2 \theta}$ is the complete elliptic integral of the first kind with modulus $k = \sqrt{u/(1-2u)}$. The series (\ref{eq19}) of ``subtracted'' $l$-modes, $h_{uu}^{(l) \, {\rm subtracted}} := h_{uu}^{(l)} - D_0$, converges like $\underset{l}{\sum} 1/l^2$.

As already mentioned, we find more convenient to work with the double contraction $h_{kk} := h_{\mu\nu} k^{\mu} k^{\nu}$ with the helical Killing vector $k^{\mu} \partial_{\mu} = \partial_t + \Omega \partial_{\varphi}$ (such that $k^{\mu} = (ds_1 / dt) u_1^{\mu}$ with $ds_1 / dt = \sqrt{1-3u}$), i.e.
\be
\label{eq22}
h_{kk}^R = (1-3u) h_{uu}^R = \sum_l \left( \left( \sum_m h_{kk}^{(l,m)} \right) - \tilde D_0 \right) ,
\ee
with a renormalized subtraction constant $\tilde D_0 := (1-3u) D_0$.

 Let us note for completeness the relation
 between $ h_{kk}^{R}(u) $ and  other versions of  Detweiler's redshift variable, such as
\be
\label{huu}
h_{uu}^{R} := h_{\mu\nu}^{R} u_1^{\mu} u_1^{\nu} = \frac{1}{1-3u}  h_{kk}^{R} ,
\ee
 or
\be
\label{u1t}
u_{(1GSF)}^t= \frac12 \frac{1}{\sqrt{1-3u}}  h_{uu}^{R} =   \frac12 \frac{1}{(1-3u)^{3/2}}  h_{kk}^{R}  .
\ee
The argument $u$ in $ h_{kk}^{R}(u) $ or $1-3u$  should not be confused with the notation $u_1^{\mu}$ used for the 4-velocity of the small particle labelled 1. [In turn
the particle label 1 should not be confused with the use of 1 as a label for 1GSF contributions. To avoid such a confusion
we have suppressed the particle label on $u^t$ in the last equation.]

In previous works \cite{Detweiler:2008ft,Sago:2008id,Blanchet:2009sd,Akcay:2012ea,Shah:2013uya}, $h_{uu}^R$ was evaluated {\it numerically}  along a (discrete) sequence of circular orbits [parametrized by $r$, $u$ or $x=(GM\Omega / c^3)^{2/3} = u + O(\nu)$].  

On the analytical side,  $h_{uu}^R$ (or equivalent functions) was evaluated by means of usual PN theory  at
increasing levels of accuracy:  2PN  \cite{Detweiler:2008ft}, and 3PN \cite{Blanchet:2009sd}.
The logarithmic contributions
to  $h_{uu}^R$ were evaluated  at 4PN  in \cite{Damour:2009sm,Blanchet:2010zd},  and even at 5PN in  \cite{Blanchet:2010zd}. In our previous work \cite{Bini:2013zaa} we showed how to analytically compute the PN expansion of
$h_{kk}^R = (1-3u) h_{uu}^R$ by using black hole perturbation theory. Such an approach avoids the subtleties 
linked to the well-known  breakdown of the usual post-Newtonian expansion arising at the 
4PN level \cite{Blanchet:1987wq}. More precisely, our approach is based on computing  the {\it post-Minkowskian} 
expansion of the function $h_{kk}^R (u)$ (weak-field expansion in powers of $G$, or equivalently in powers of 
$u = GM/c^2 r$), by using the version of  
Regge-Wheeler-Zerilli (RWZ) black hole perturbation theory that has been developed by 
Mano, Suzuki and Takasugi (MST) \cite{Mano:1996vt,Mano:1996mf,Mano:1996gn}.
[Let us note that Refs.  \cite{Sago:2002fe,Nakano:2003he,Hikida:2004jw,Hikida:2004hs} made valient attempts to apply the MST formalism 
 to GSF theory, but were bogged down by gauge-dependent issues. The fact that
we consider a gauge-invariant quantity allows us to get novel results.] 
In our short report  \cite{Bini:2013zaa},  we only gave the result of our analysis at the 4PN level, by providing the analytical
expression of the non-logarithmic coefficient entering   $h_{kk}^R (u)$ at 4PN (i.e. the coefficient of $u^5$).
In the present paper, we report on our higher-PN results: at 5PN, 5.5PN and 6PN.

\subsection{Solving in terms of their sources the radial functions $R_{lm\omega }^{\rm (odd)}$, $R_{lm\omega }^{\rm (even)}$ and $Z_{lm\omega} (r)$  entering  the Regge-Wheeler-Zerilli metric perturbation}

In the series (\ref{eq22}) over $l$, the low-multipole contributions $l=0$ and $l=1$ (even and odd) can be computed from the corresponding exact results of Zerilli \cite{Zerilli:1971wd}. Note the subtlety that we are computing here $h_{kk}^R (u)$ in
an {\it asymptotically flat} gauge. One must accordingly rephrase the classic results of  Zerilli \cite{Zerilli:1971wd}. We recall
in this respect that the $l=0$ contribution to the Lorenz-gauge metric perturbation fails to be (manifestly) asymptotically
flat  \cite{Detweiler:2003ci,Barack:2005nr}.  When using the Lorenz gauge, one must accordingly introduce a correction
to compute  flat-gauge quantities \cite{Sago:2008id,Damour:2009sm,Akcay:2012ea}.

On the other hand, the ``dynamical'' multipoles of order $l \geq 2$ are more difficult to evaluate. We started from the {\it corrected} version of the RWZ equations derived by Sago, Nakano and Sasaki \cite{Sago:2002fe}. [Beware, however, that the sign of the source term $B_{lm}^{(0)}$ in Table~I there should be $-i(\ldots)$ instead of $+i(\ldots)$; the later papers \cite{Nakano:2003he,Hikida:2004jw} give the correct sign.] 

The original RWZ formalism expresses an {\it odd-parity} metric perturbation $h_{\mu\nu}^{(l,m)}$, with frequency $\omega$, in terms of a radial function $R_{lm\omega}^{({\rm odd})} (r)$ satisfying a Regge-Wheeler(RW)-type equation
\be
\label{eq23}
{\mathcal L}^{(r)}_{\rm (RW)}[R_{lm\omega }^{\rm (odd)}]=S_{lm\omega}^{\rm (odd)}(r) \, .
\ee
Here ${\mathcal L}^{(r)}_{\rm (RW)}$ denotes the RW operator
\begin{eqnarray}
\label{eq24}
{\mathcal L}^{(r)}_{\rm (RW)}&=&\left(1-\frac{2M}{r}  \right)^2 \frac{d^2}{dr^2} +\frac{2M}{r^2}\left(1-\frac{2M}{r}  \right) \frac{d}{dr }   \nonumber \\
&+&[\omega^2 -V_{\rm (RW)}(r)]\nonumber\\
&=& \frac{d^2}{dr_*^2}   +[\omega^2 -V_{\rm (RW)}(r)]
\end{eqnarray}
with $d/dr_* = f(r) d/dr$ (with $f(r) :=1-2M/r$), and a RW potential
\be
\label{eq25}
V_{\rm (RW)}(r)=\left(1-\frac{2M}{r}  \right) \left(\frac{l(l+1)}{r^2}-\frac{6M}{r^3}  \right)\, .
\ee 
Note that we denote here the large background mass (which, strictly speaking, should be $m_2$) by $M$.
This is allowed in our calculation which is to linear order in $m_1/m_2$.  (Here, as often in the following, we omit to include a label $l\omega$ indicating the $l\omega$-dependence of various objects.)
The odd-parity source term in Eq.~(\ref{eq23}) (given by Eq.~(A35) of \cite{Sago:2002fe}) is of the form
\be
\label{eq26}
S_{lm\omega}^{\rm (odd)}(r)=s^{\rm (o)}_0\delta(r-r_0)+s^{\rm (o)}_1\delta'(r-r_0)\,,
\ee
where $r_0$ denotes the radius of the circular orbit of particle 1. 

On the other hand, the original RWZ formalism expresses a (monochromatic) {\it even-parity} $h_{\mu\nu}^{(l,m)}$ in terms of a radial function $Z_{lm\omega} (r)$ satisfying a Zerilli-type equation \cite{Zerilli:1971wd}, involving a more complicated potential than the RW equation (\ref{eq23}). Using a result of Chandrasekhar \cite{Chandrasekhar:1975}, one can, however, replace the pair ${\mathcal Z}_{lm\omega} := (Z_{lm\omega} (r) , dZ_{lm\omega} (r)/dr)$ by a new pair of functions, say ${\mathcal R}_{lm\omega} := (R_{lm\omega}^{({\rm even})} (r) , dR_{lm\omega}^{({\rm even})} (r)/dr)$, satisfying a simpler RW-type equation, say
\be
\label{eq27}
{\mathcal L}^{(r)}_{\rm (RW)}[R_{lm\omega}^{\rm (even)}]=S_{lm\omega}^{\rm (even)}(r)\,.
\ee

The price to pay for this simplification of the potential is: (i) the transformation between ${\mathcal Z}_{lm\omega}$ and ${\mathcal R}_{lm\omega}$ involves source terms, and (ii) the new even-parity source term is connected to the original Zerilli one by an expression of the form $S_{lm\omega}^{\rm (even)}={\mathcal A}_{11} (r) S_{lm\omega}^{\rm (Z)}+{\mathcal A}_{12}(r) \frac{d}{dr} (S_{lm\omega}^{\rm (Z)})$. As a consequence the new even-parity source term in Eq.~(\ref{eq27}) is of the form
\be
\label{eq28}
S_{lm\omega}^{\rm (even)} =  s^{\rm (e)}_0\delta(r-r_0)+s^{\rm (e)}_1\delta'(r-r_0)+s^{\rm (e)}_2\delta''(r-r_0) \, .
\ee

At this stage, the problem is reduced to solving some RW equation (one for each $lm\omega$ and each parity) with given (distributional) source terms. The source terms derive from the spherical harmonics projection of the distributional stress-energy tensor of particle 1:
\be
\label{eq29}
T_1^{\mu\nu} (x^{\lambda}) = m_1 (-g)^{-1/2} \int ds_1 u_1^{\mu} u_1^{\nu} \delta^{(4)} (x^{\lambda} - y_1^{\lambda} (s_1)) \, .
\ee
As a consequence, in the case of a circular orbit, the (discrete) frequencies entering $T_1^{\mu\nu} (x)$ (and therefore $h_{\mu\nu} (x)$) are related to the basic orbital frequency $\Omega = d\phi_1 / dt_1$ and the ``magnetic'' number $m$ by $\omega = \omega_m := m\Omega$.

The solution of the RW equations (\ref{eq23}), (\ref{eq27}) is determined by the choice of the boundary conditions incorporated in a Green's function $G(r,r')$  normalized so as to satisfy
\be
\label{eq30}
{\mathcal L}^{(r)}_{\rm (RW)} G(r,r')=f(r')\delta (r-r')\,.
\ee
Such a Green's function can be generally expressed in terms of two, specially chosen, independent {\it homogeneous }solutions of the RW operator (and of the Heaviside step function $H(x)$)
\begin{eqnarray}
\label{eq31}
G(r,r')&=&\frac{1}{W}\Bigl[X_{\rm (in)}(r)X_{\rm (up)}(r')H(r'-r) \nonumber \\
&+&X_{\rm (in)}(r')X_{\rm (up)}(r)H(r-r')  \Bigl] \\
&\equiv & G_{\rm (in)}(r,r')H(r'-r)+G_{\rm (up)}(r,r')H(r-r') \nonumber \,,
\end{eqnarray}
where 
\be
{\mathcal L}_{({\rm RW})}^{(r)} X_{({\rm in})} (r) = 0 = {\mathcal L}_{({\rm RW})}^{(r)} X_{({\rm up})} (r) \, , 
\ee
and where $W$ denotes the Wronskian
\begin{eqnarray}
\label{eq32}
W&=&\left(1-\frac{2M}{r}  \right)\biggl[X_{\rm (in)}(r)\frac{d}{dr }X_{\rm (up)}(r )\nonumber \\
&-&\frac{d}{dr}X_{\rm (in)}(r)X_{\rm (up)}(r) \biggl]={\rm constant} \,.
\end{eqnarray}
The physical Green's function we are interested in is the {\it retarded} one. It is obtained, as usual, by choosing for $X_{({\rm in})}^{l\omega}$ a solution that is incoming from $r = +\infty$ (and purely ingoing on the horizon), and for $X_{({\rm up})}^{l\omega}$ a solution that is upgoing from the horizon (and purely outgoing at infinity). This uniquely determines the solutions of the {\it inhomogeneous} even-parity and odd-parity RW equations, namely
\be
\label{eq33}
R_{lm\omega}^{\rm (even/odd)}(r)=\int dr' G(r,r')f(r')^{-1}S_{lm\omega}^{\rm (even/odd)}(r')\,.
\ee

Note that the distributional nature of the radial source functions, notably $S_{lm\omega}^{\rm even} (r) \ni \delta'' (r-r_0)$, implies that, e.g., $R_{lm\omega}^{\rm even} (r)$ is not only discontinuous as $r$ crosses $r_0$, but (formally) contains a contribution $\propto \delta (r-r_0)$.

\subsection{Evaluating the even and odd multipolar pieces  $h_{kk,lm}^{\rm (even/odd)}$ of $h_{kk}$ in terms of homogeneous solutions $X_{l\omega}^{\rm (in)}$, $\ X_{l\omega}^{\rm (up)}$ of the Regge-Wheeler equation}

Having determined $R_{lm\omega}^{({\rm even}/{\rm odd})} (r)$ by the (distributional) formula (\ref{eq33}), one can then compute the original Zerilli radial functions $(Z_{lm\omega} (r) , dZ_{lm\omega} (r) / dr)$, and thereby evaluate the metric perturbation $h_{\mu\nu}^{(lm\omega)} (r)$, and, in particular, the double contraction
\begin{eqnarray}
\label{eq34}
h_{kk}(t,r,\theta,\phi)&= &h^{\rm (RWZ)}_{tt}+2\Omega h^{\rm (RWZ)}_{t\phi}+\Omega^2 h^{\rm (RWZ)}_{\phi\phi} \nonumber \\
&=&\sum_{l, m}e^{-i\omega_m t}  h_{kk}{}_{ l m  \omega }^{\rm (RWZ)}
\end{eqnarray}
we are interested in. Here, $h_{kk} (x^{\lambda})$ is evaluated at a generic field point $x^{\lambda} = (t,r,\theta,\phi)$. The next step is to restrict $t, \theta$ and $\phi$ to the values corresponding to the considered instantaneous position of particle~1, say $t,\theta = \pi / 2$ and $\phi = \phi_1 (t) = \Omega t$ (in the equatorial plane of the background Schwarzschild metric). [This eliminates the apparent $t$-dependence of $h_{kk}$ in Eq.~(\ref{eq34}).] At this stage, $h_{kk}$ depends only on $r$. Considering the two limits $r \to r_0^-$ and $r \to r_0^+$, we have checked that they yield the same result for the value of $h_{kk}$ at the location $r=r_0$ of particle~1. [This confirms the idea of Detweiler \cite{Detweiler:2008ft} that the gauge-invariant quantity $h_{uu}^R$ can be correctly evaluated on the world line of $y_1$ even if one uses a gauge (such as the RWZ one) where $h_{\mu\nu} (x;y_1)$ has a worse behaviour than its Lorenz-gauge version.]

Our final result for $h_{kk}^{(l,m)} \equiv h_{kk}^{(l,m)} (r_0)$, which enters Eq.~(\ref{eq22}), is the sum of an even and an odd contribution
\be
h_{kk}^{(l,m)}= h_{kk,lm}^{\rm (even)} + h_{kk,lm}^{\rm (odd)}
\ee
The odd contribution takes the rather simple form
\begin{eqnarray}
\label{eq35odd}
h_{kk,lm}^{\rm (odd)}&=&-|\partial_\theta Y_{lm}(\pi/2,0)|^2 \frac{8 \pi \Gamma}{r_0^3W\Lambda} M f_0^2 \times\nonumber \\
&&\times \left[r_0 \frac{d X_{l\omega}^{\rm (in)}}{dr_0}+X_{l\omega}^{\rm (in)}\right] 
\left[r_0 \frac{d X_{l\omega}^{\rm (up)}}{dr_0}+X_{l\omega}^{\rm (up)}\right]\,. \nonumber \\
\end{eqnarray}
Here $\Gamma := u_1^t = (1-3u)^{-1/2}$, and $\Lambda := \frac14 (l-1) l(l+1)(l+2)$. 

The corresponding result for $h_{kk,lm}^{({\rm even})}$ is more complicated, and reads
\begin{eqnarray}
\label{eq35even}
h_{kk,lm}^{\rm (even)}&=&- |Y_{lm}(\pi/2,0)|^2 \frac{8\pi \Gamma  }{r_0^5 \Lambda W (r_0^3\Lambda^2 +9M^3m^2)}\times \nonumber \\
&&\times {\mathcal F}(X^{\rm (in)};r_0){\mathcal F}(X^{\rm (up)};r_0)
\end{eqnarray}
where
\be
{\mathcal F}(X;r_0)=\tilde  \alpha(r_0) \frac{dX(r_0)}{dr_0}+\tilde \beta(r_0)X(r_0)
\ee
with (using the notation $\lambda := (l-1)(l+2)/2$ such that $\Lambda=\lambda(\lambda+1)$)
\begin{eqnarray}
\tilde \alpha(r_0)&=& r_0^2 f_0 [\Lambda r_0^2 f_0+3M^2 (\lambda+1-m^2)]\nonumber\\
\tilde \beta(r_0)&=& r_0^2(r_0-M)\lambda^3+r_0[6M^2-r_0(5M-2r_0 \nonumber\\
&&+ \, Mm^2)]\lambda^2 \nonumber\\
&+&(r_0^3-4Mr_0^2-Mr_0^2m^2+9M^2r_0-6M^3)\lambda \nonumber \\
&&- \, 3Mr_0f_0(m^2-1)\,.
\end{eqnarray}

Note that both type of expressions have  the usual ``one-loop'' structure: $({\rm source})^* \times (\mbox{propagator}) \times ({\rm source})$ (where the star indicates taking a complex conjugate, as appropriate for a transition amplitude).
  For the even case, the source term is proportional  to $Y_{lm} (\pi/2,0)$,
while, for the odd case,  the source term is proportional  to $\partial_\theta Y_{lm} (\pi/2,0)$.  In both cases the propagator
has the characteristic Green's function structure, i.e. it is bilinear in the {\it homogeneous} RW solutions
$X_{l\omega}^{\rm (in)}$ and $ X_{l\omega}^{\rm (up)}$,
or their derivatives.

\subsection{ Strategy for analytically computing the PN expansion of   $h_{kk}^R(u)$ }

To evaluate $h_{kk}^R$ from the RWZ results (\ref{eq35odd}), (\ref{eq35even})  one still needs, according to Eq.~(\ref{eq22}), to: (i) sum over $m$ from  $-l$ to $+l$; (ii) subtract $\tilde D_0$; and, finally, (iii) sum the result of (i) and (ii) over $\ell = 0,1,2,\ldots$. Even the first (finite) sum over $m$ is quite nontrivial to compute analytically because one must remember that the index $\omega$ on the two solutions $X_{l\omega}^{({\rm in})} , X_{l\omega}^{({\rm up})}$ entering (\ref{eq35odd}), (\ref{eq35even})  actually refers to $\omega_m = m\Omega$, not to mention the fact that one needs to obtain explicit, analytic expressions for the two homogeneous solutions $X_{l\omega}^{({\rm in})} (r)$ and $X_{l\omega}^{({\rm up})} (r)$. The latter problem has been formally solved by Mano et al. \cite{Mano:1996vt,Mano:1996mf,Mano:1996gn} who gave analytic expressions for $X_{l\omega}^{({\rm in})}$ and $X_{l\omega}^{({\rm up})}$ in the form of series of hypergeometric functions (of the usual, Gauss, type 
for $X^{({\rm in})}$ and of the confluent type for $X^{({\rm up})}$). More precisely, they obtained expressions of the type
\begin{eqnarray}
\label{eq36}
X_{l\omega}^{({\rm in})} (r) &= &C_{({\rm in})}^{\nu} (x) \sum_{n=-\infty}^{+\infty} a_n^{\nu} \times \\
&&\times \overline F (n+\nu -1 -i\epsilon, -n-\nu-2-i\epsilon, 1-2i \epsilon ; x ] \, ,  \nonumber
\end{eqnarray}
\begin{eqnarray}
\label{eq37}
X_{l\omega}^{({\rm up})} (r) &= &C_{({\rm up})}^{\nu} (z) \sum_{n=-\infty}^{+\infty} a_n^{\nu} (-2iz)^n \times \\
&&\times \overline\Psi (n+\nu +1 -i\epsilon, 2n+2\nu+2;-2iz) \, . \nonumber
\end{eqnarray}
Here, $x=1-c^2 r/2GM$, $z = \omega r/c$, $\epsilon = 2GM\omega / c^3 = 2mGM\Omega/c^3$
\begin{eqnarray}
\label{eq38}
\nu&=&l+\frac{1}{2l+1} \biggl[-2 -\frac{s^2}{l(l+1)} \nonumber \\
&&+\frac{[(l+1)^2-s^2]^2}{(2l+1)(2l+2)(2l+3)} \\
&&-\frac{(l^2-s^2)^2}{(2l-1)2l(2l+1)}\biggl]\epsilon^2 + O(\epsilon^4)\,, \nonumber
\end{eqnarray} 
is an $\epsilon$-modified avatar of $l$ \cite{Mano:1996vt,Mano:1996mf,Mano:1996gn}  (with $s^2=4$ in the present spin~2 case; see appendix A for more details), and
$$
C_{({\rm in})}^{\nu} (x) = c_{({\rm in})} e^{i\epsilon [(x-1)-\ln(-x)]} (1-x)^{-1} \, ,
$$
$$
C_{({\rm up})}^{\nu} (z) = c_{({\rm up})} e^{iz}z^{\nu+1}\left( 1-\frac{\epsilon}{z} \right)^{-i\epsilon}2^\nu e^{-\pi \epsilon}e^{-i\pi (\nu+1)}\,,
$$
$$
\overline F (a,b,c;x) = \frac{\Gamma (a) \Gamma(b)}{\Gamma (c)} \quad F(a,b,c;x) \, ,
$$
$$
\overline\Psi (a,b;\zeta) = \frac{\Gamma (a-2) \Gamma (a)}{\Gamma (a^*) \Gamma (a^*+2)} \Psi (a,b;\zeta)
$$
(with $a^*$ denoting the complex conjugate of $a$; and $\Psi$ the second Kummer function).  
Here the quantities $c_{({\rm in})}=\eta^{\alpha_{\rm (in)}^{(l)}}$ and $c_{({\rm up})}=\eta^{\alpha_{\rm (up)}^{(l)}}$ are some $l$-dependent powers of $\eta:=1/c$ that are defined so that $C_{({\rm in})}^{\nu} (x)$ and $C_{({\rm up})}^{\nu} (z)$ both start with zeroth order in $\eta$.
For example, for $l=2$ we take $c_{({\rm in})}=\eta^{-2}$ and $c_{({\rm up})}=\eta^{-3}$ so that
\begin{eqnarray*}
C_{({\rm in})}(r)&=&\frac{2M}{r}-2i M\omega \eta -Mr\omega^2 \eta^2 +O(\eta^3)\nonumber\\
C_{({\rm up})}(r)&=&-4\omega^3 r^3 -4i \omega^4 r^4\eta+2 \omega^5 r^5\eta^2 +O(\eta^3)\,.
\end{eqnarray*}

Finally, the two-sided sequence of coefficients $a_n^{\nu}$ entering both series (\ref{eq36}) and (\ref{eq37}) are obtained by solving a three-term recursion relation, $\alpha_n^{\nu} a_{n+1}^{\nu} + \beta_n^{\nu} a_n^{\nu} + \gamma_n^{\nu} a_{n-1}^{\nu} = 0$, obtained by Mano et al. (see Eqs.~(2.5)--(2.8) in \cite{Mano:1996mf}).  We have solved this three-term relation between $n=-11$
and $n=+13$ (included) by initiating the recursion with $a_{13}=0$ and $a_{-11}=0$.  
This is sufficient for getting the crucial
hypergeometric-expanded  $X_{l\omega}^{({\rm up})} $  to the 6PN level (see below).

In principle, the expressions (\ref{eq36}), (\ref{eq37}) can be used to compute $h_{kk}^{(l,m)} (r_0)$ for all values of $l$ and $m$.  However, it would be difficult to use them to treat analytically the case of arbitrarily high values of $l$ (as needed for
analytically implementing the mode-sum regularization procedure). Fortunately,
for each pre-decided PN accuracy of the final result, we do not need to use the full power of the 
hypergeometric series expansions (\ref{eq36}), (\ref{eq37}). For instance, if we are interested in having 4PN
accuracy for $h_{kk}^R$,
 the work of Ref.~\cite{Blanchet:1987wq} has shown that, at the 4PN level, the subtle (formally infra-red divergent) mixing of near-zone and wave-zone effects only occurs through mass-type quadrupolar $(l=2)$ couplings. This indicates that,
 at the 4PN level,  the full power of the hypergeometric  series expansions is only needed to correctly get the $l=2$ contribution to $h_{kk}^{(l,m)}$, and that a usual PN expansion of the RWZ equations is accurate enough to evaluate the $l \geq 3$ contributions. On the other hand, if one aims at a higher PN accuracy, one needs to apply the  MST technology to correspondingly higher values of $l$.  Actually,  the main role of the MST expansions is to correctly capture the transition
 between the near-zone and the wave-zone, i.e., within our context,
  to correctly incorporate all tail effects in the near-zone metric perturbation
 $ h_{\mu\nu} (x^{\lambda})$ that we use to compute $h_{kk}(x)$.  On the one hand, 
 we know \cite{Blanchet:1987wq,Blanchet:1992br}   that 
 these tail effects enter the near-zone metric as hereditary modifications of radiation reaction.  On the other hand,
 we know that the PN order at which (leading-order) radiation effects enter the dynamics increases with
 the multipolarity $l$, and also depends on the even/odd character of the considered multipole. More precisely,
 from the study of radiation reaction linked  to higher multipoles in Ref. \cite{Blanchet:1984wm},
 one knows that mass-type (i.e. even) multipoles of degree $l$ enter radiation reaction at the PN order $1/c^{2 l+1}$,
 while spin-type (i.e. odd)  multipoles of degree $l$ enter radiation reaction at the PN order $1/c^{2 l+3}$. 
 Tail effects linked to the total mass of the system then modify in an hereditary way these leading-order effects by adding corrections proportional to $GM/c^3$ and its higher powers. In other words, the IR-delicate hereditary effects (for which we need
 MST technology) arise, from each multipole, at the following PN orders: (i) from $l=2$ even, at orders 
 $1/c^5 \times (1/c^3+1/c^6+1/c^9+\cdots)  $ (i.e. 4PN, 5.5PN, 7PN, etc); (ii) from $l=2$ odd, at  orders $1/c^7 \times (1/c^3+1/c^6+1/c^9+\cdots) $ (i.e. 5PN, 6.5PN, 8PN, etc); (ii) from $l=3$ even, at  orders $1/c^7 \times (1/c^3+1/c^6+1/c^9+\cdots)  $  (i.e. 5PN, 6.5PN, 8PN, etc); (iii)   from $l=3$ odd, at  orders $1/c^9 \times (1/c^3+1/c^6+1/c^9+\cdots)  $  (i.e. 6PN, 7.5PN, 9PN, etc);  from $l=4$ even, at  orders $1/c^9 \times (1/c^3+1/c^6+1/c^9+\cdots)  $  (i.e. 6PN, 7.5PN, 9PN, etc);  
 and so on.  As our aim is to reach the 6PN accuracy, we see that it is sufficient to use hypergeometric series up to 
 $l=4$ (even).  For higher values of  $l$ (and, actually, also for the odd, $l=4$ contribution) it must be sufficient to use
 PN-type solutions of the RW equations for $X$.  In addition,  the reasoning above shows that hypergeometric
 expansions should only be needed to get right the ``up'' part of the Green's function, which is the part which incorporates
 the transition between the near zone and the wave zone. [The ``in'' part of the  Green's function, related to the matching
 between the near zone and the horizon, is not expected to influence the near zone metric until  4PN levels beyond
 radiation reaction \cite{Poisson:1994yf}, i.e. terms of order $1/c^5 \times 1/c^8 = 1/c^{13}$ (see Sec. 3.4 of Ref.~\cite{Sasaki:2003xr});
 moreover these terms are of radiative nature, and will start modifying the conservative effects we are considering
 only at a strictly higher order, i.e. strictly beyond the 6.5PN level.]
 We have explicitly checked the correctness of this expectation by studying in detail the difference between the exact expansions (\ref{eq36}), (\ref{eq37}) and their PN counterparts.

 \subsection{Expanding the hypergeometric-series ``in'' and ``up'' homogeneous solutions in powers of  $\eta := 1/c$ }
 
 In this work we shall denote the basic PN expansion parameter as  $\eta := 1/c$. [We use the Greek letter   $\eta$ because
 we have followed MST and other Japanese authors in using  $\epsilon$ to denote  $ 2GM\omega / c^3$. Note that 
 $\epsilon$ is of PN order $\eta^3$.]   The PN expansion parameter   $\eta := 1/c$ has a double meaning: it keeps track
 both of the near-zone expansion (in powers of $  \omega r/c$), and of the weak-field expansion (in powers of $GM/(r c^2)$).
 The  expansions of  $X_{l\omega}^{\rm in}$ and $X_{l\omega}^{\rm up} $ in  powers of $\eta$
 will have the general form
  \begin{eqnarray}
\label{eq41a}
X_{l\omega}^{{\rm in} } (r) &= &  \tilde c_{\rm in}  r^{l+1}  \sum_{k=0}^{k_{\rm max}} A_k^{{\rm in} (l)} \eta^k  \, ,
\end{eqnarray}
\begin{eqnarray}
\label{eq42a}
X_{l\omega}^{{\rm up} } (r) &= & \tilde c_{\rm up}  r^{-l}  \sum_{k=0}^{k_{\rm max}} A_k^{{\rm up} (l)} \eta^k  \, ,
\end{eqnarray}
 where  $\tilde c_{\rm in}$ and $\tilde c_{\rm up}$ are some normalization coefficients, where  $A_0^{{\rm in} (l)}=1=A_0^{{\rm up} (l)}$,
and where the logarithmic dependence in $\eta$  is 
absorbed in the expansion coefficients  $A_k$.  We shall consider the  $A_k$'s as functions of the two 
quantities $X_1 = GM/r$ (linked to the weak-field expansion), and  $\sqrt{X_2} := \omega r$  (linked to
the near-zone expansion).  
[The auxiliary  arguments $X_1, X_2$ should not be confused with the
 dependent variables $X_{l\omega}^{{\rm in} }, X_{l\omega}^{{\rm up} }$.]
Modulo some logarithmic 
dependence in $r$,  the  $A_k$'s are polynomials in $X_1$ and  $\sqrt{X_2}$. 

 Let us indicate the structure of the $\eta$ expansion of the crucial  hypergeometric {\it up} solutions (\ref{eq37}) for
 the particular case $l=2$.
 We derived similar explicit results for  $l= 3$ and $4$.  In all cases we have extended the expansion up to 
 $k_{\rm max} =12$, corresponding to the (fractional) 6PN level.
 When using $c_{\rm up} = c^3 = \eta^{-3}$ in the definition of $C_{({\rm up})}^{\nu} (z)$ in Eq.  (\ref{eq37}) , 
  the $\eta$ expansion of the hypergeometric ``up'' solution for $l=2$, $X_{\rm (up)}^{HG, l=2}$, reads
\be
\label{eq39}
X_{\rm (up)}^{HG, l=2}= -\frac{i}{16\omega^2 r^2} \sum_{k=0}^{12} A_k^{{\rm up} (HG,l=2)} \eta^k
\ee
where HG stands for \lq\lq hypergeometric" and 
the explicit values of the coefficients $A_k^{\rm up}$  read
\begin{eqnarray}
\label{eq40}
A_0^{{\rm up}\, (HG,\, l=2)}&=& 1 \, ,  \qquad
A_1^{{\rm up}\, (HG,\, l=2)}= 0 \, ,
\nonumber\\
A_2^{{\rm up}\, (HG,\, l=2)}&=& \frac16 X_2+\frac53 X_1  \, ,\nonumber\\
\end{eqnarray}
\begin{widetext}
\begin{eqnarray}
\label{Akup}
A_3^{{\rm up}\, (HG,\, l=2)}&=& \left(6i\gamma-2\pi-\frac{43}{6}i\right)X_1 \sqrt{X_2}\nonumber\\
A_4^{{\rm up}\, (HG,\, l=2)}&=& \frac76 X_1X_2+\frac{1}{24}X_2^2+\frac{20}{7}X_1^2 \nonumber\\
A_5^{{\rm up}\, (HG,\, l=2)}&=& \biggl[ \left(10i\gamma-\frac{10}{3}\pi-\frac{215}{18}i\right)X_1^2 
+\left(-\frac{43}{36}i+i\gamma -\frac13 \pi\right) X_1X_2 
 +\frac{1}{45}iX_2^2 \biggl]\sqrt{X_2}\nonumber\\
A_6^{{\rm up}\, (HG,\, l=2)}&=&  5 X_1^3+\biggl(-18\gamma^2+\frac{7}{3}\pi^2-12i\pi\gamma 
 -\frac{272551}{8820}+\frac{214}{105} \ln(2\omega r \eta)+\frac{4729}{105}\gamma \nonumber\\
&& +\frac{43}{3}i\pi\biggl)X_1^2 X_2
+\frac{7}{24} X_1 X_2^2-\frac{1}{144}X_2^3 \nonumber\\
A_7^{{\rm up}\, (HG,\, l=2)}&=&\biggl[ \left(-\frac{430}{21} i+\frac{120}{7}i\gamma-\frac{40}{7}\pi\right) X_1^3 
+\left(-\frac{301}{36} i-\frac{7}{3}\pi+7i\gamma\right) X_2X_1^2\nonumber\\
&& +\left(-\frac{1}{12}\pi-\frac{43}{144}i+\frac{1}{4}i\gamma\right) X_2^2 X_1 
-\frac{1}{630}iX_2^3 \biggl]\sqrt{X_2} \nonumber\\
A_8^{{\rm up}\, (HG,\, l=2)}&=& \frac{80}{9} X_1^4+\biggl(-20i \pi\gamma+\frac{215}{9}i\pi+\frac{4729}{63}\gamma 
-30\gamma^2+\frac{35}{9}\pi^2-\frac{1477681}{26460} \nonumber\\
&&+\frac{214}{63}\ln(2\omega r \eta)\biggl) X_2X_1^3
 +\biggl(-\frac{76169}{17640}+\frac{7}{18}\pi^2+\frac{43}{18}i \pi-2i \pi\gamma \nonumber\\
&& +\frac{4729}{630}\gamma-3\gamma^2+\frac{107}{315}\ln(2\omega r \eta)\biggl)X_2^2X_1^2
+\left(\frac{3547}{10800}-\frac{4}{45}\ln(2\omega r \eta)-\frac{2}{9}\gamma\right) X_2^3X_1 
+\frac{1}{3456} X_2^4 
\nonumber\\
A_9^{{\rm up}\, (HG,\, l=2)}&=&\left[ \left(-10\pi-\frac{215}{6}i+30i\gamma\right) X_1^4 \right.
+\left(-2\pi^3-36i\gamma^3+\frac{4943}{35}i\gamma^2 \right. 
+\frac{428}{35}i\gamma \ln(2\omega r \eta)\nonumber\\
&&+\frac{1691881}{17640}i-8i\zeta(3)+14i\pi^2\gamma
+\frac{272551}{4410}\pi-\frac{4601}{315}i \ln(2\omega r \eta)
-\frac{428}{105}\pi \ln(2\omega r \eta) +36\gamma^2\pi
\nonumber\\&&
\left.-\frac{9458}{105}\pi\gamma-\frac{882067}{4410}i \gamma-\frac{9251}{630}i\pi^2\right) X_2X_1^3
+\left(\frac{7}{4}i\gamma-\frac{7}{12}\pi-\frac{301}{144}i\right) X_2^2X_1^2 
\nonumber \\
&&
+\left(\frac{121}{3360}i-\frac{1}{24} i \gamma+\frac{1}{72}\pi \right)X_2^3 X_1 
\left.+\frac{1}{22680}i X_2^4 \right]\sqrt{X_2}  
\end{eqnarray}
\begin{eqnarray}
\label{Akup_2}
A_{10}^{{\rm up}\, (HG,\, l=2)}&=& 
16 X_1^5+\Biggl(\frac{20}{3}\pi^2-\frac{240}{7}i\pi\gamma+\frac{860}{21}i\pi 
-\frac{220798}{2205}-\frac{360}{7}\gamma^2+\frac{856}{147} \ln(2\omega r \eta) 
+\frac{18916}{147}\gamma\Biggl) X_2X_1^4
\nonumber\\&&
+\Biggl(\frac{107}{45} \ln(2\omega r \eta) -21\gamma^2+\frac{301}{18}i\pi 
+\frac{4729}{90}\gamma-14i\pi\gamma-\frac{33862}{945}+\frac{49}{18}\pi^2\Biggl) X_2^2X_1^3\nonumber\\
&&+\Biggl(-\frac{45901}{70560}+\frac{43}{72}i\pi+\frac{107}{1260} \ln(2\omega r \eta) 
+\frac{4729}{2520}\gamma-\frac{1}{2}i\pi\gamma-\frac{3}{4}\gamma^2+\frac{7}{72}\pi^2\Biggl) X_2^3X_1^2\nonumber\\
&&+\left(\frac{1}{63}\gamma-\frac{108721}{4233600}+\frac{2}{315} \ln(2\omega r \eta) \right) X_2^4X_1
-\frac{1}{172800} X_2^5 \nonumber\\
A_{11}^{{\rm up}\, (HG,\, l=2)}&=& \Biggl[ \left(\frac{160}{3}i\gamma-\frac{1720}{27}i-\frac{160}{9}\pi\right) X_1^5
+\Biggl(\frac{428}{21}i\gamma \ln(2\omega r \eta) -60i\gamma^3+60\gamma^2\pi 
+\frac{1477681}{13230}\pi+\frac{70}{3}i\pi^2\gamma-\frac{10}{3}\pi^3\nonumber \\
&&-\frac{428}{63}\pi \ln(2\omega r \eta) -\frac{9458}{63}\pi\gamma
+\frac{4943}{21}i\gamma^2-\frac{9251}{378}i\pi^2-\frac{40}{3}i\zeta(3) 
+\frac{30320033}{158760}i-\frac{4601}{189} i \ln(2\omega r \eta) \nonumber \\
&&-\frac{4755113}{13230}i\gamma \Biggl) X_2X_1^4 
+\Biggl(\frac{76169}{8820}\pi+\frac{3181751}{317520}i-\frac{4729}{315}\pi\gamma+6\gamma^2\pi
-\frac13 \pi^3-6i\gamma^3-\frac{149987}{5292}i\gamma-\frac43 i\zeta(3)\nonumber \\
&&+\frac{214}{105}i \gamma \ln(2\omega r \eta) 
-\frac{4601}{1890}i \ln(2\omega r \eta) +\frac{7}{3}i\pi^2\gamma+\frac{4943}{210}i\gamma^2
-\frac{214}{315}\pi \ln(2\omega r \eta) -\frac{9251}{3780}i\pi^2\Biggl) X_2^2 X_1^3\nonumber
\end{eqnarray}
\begin{eqnarray}
&&+\Biggl(\frac{8}{45}\pi \gamma+\frac{932}{1575}i \ln(2\omega r \eta) -\frac{8}{15}i\gamma \ln(2\omega r \eta)
-\frac{3532427}{2268000} i-\frac{14}{15}i\gamma^2-\frac{14501}{37800}\pi+\frac{6457}{2520}i\gamma 
\nonumber\\&&
 \left.+\frac{8}{45}\pi \ln(2\omega r \eta) -\frac{1}{135}i\pi^2\right) X_2^3X_1^2
+\left(-\frac{389}{311040} i-\frac{1}{1728}\pi+\frac{1}{576}i\gamma\right) X_2^4X_1 \nonumber \\
&&-\frac{1}{1496880}i X_2^5\Biggl]\sqrt{X_2} \nonumber\\
A_{12}^{{\rm up}\, (HG,\, l=2)}&=& 
\frac{320}{11} X_1^6+\Biggl(\frac{215}{3}i\pi-60i\pi \gamma-\frac{4790743}{26460}
+\frac{214}{21} \ln(2\omega r \eta) +\frac{35}{3}\pi^2-90\gamma^2 
+\frac{4729}{21}\gamma \Biggl) X_2 X_1^5\nonumber\\
&&+\Biggl(-\frac{420397183}{661500}\gamma-\frac{5164}{105}\zeta(3)
-\frac{7469477}{132300}\pi^2+\frac{121}{90}\pi^4+\frac{14243011}{22050}\gamma^2\nonumber \\
&&+\frac{965664058429471}{4453225938000}
-\frac{19861979}{330750} \ln(2\omega r \eta)
-\frac{10314}{35}\gamma^3+54\gamma^4+\frac{22898}{11025} \ln^2(2\omega r \eta) \nonumber\\  
&&+\frac{1012006}{11025}\gamma \ln(2\omega r \eta) +\frac{29251}{315}\pi^2\gamma 
 -\frac{1691881}{8820}i\pi+\frac{359}{35}i\pi^3-\frac{1284}{35}\gamma^2 \ln(2\omega r \eta) \nonumber \\
&&+\frac{214}{45} \ln(2\omega r \eta) \pi^2+48\zeta(3)\gamma-42\pi^2\gamma^2
-\frac{856}{35}i\pi\gamma \ln(2\omega r \eta) +\frac{882067}{2205}i\pi\gamma\nonumber\\
&&+72i\pi\gamma^3-12i\pi^3\gamma+16 i \pi\zeta(3)-\frac{9886}{35}i\gamma^2\pi
+\frac{9202}{315} i\pi \ln(2\omega r \eta)\Biggl)X_2^2X_1^4\nonumber\\
&& +\Biggl(\frac{4729}{360}\gamma+\frac{49}{72}\pi^2-\frac{100693}{12960}-\frac{7}{2}i\pi\gamma
+\frac{301}{72}i\pi+\frac{107}{180} \ln(2\omega r \eta) -\frac{21}{4}\gamma^2\Biggl) X_2^3 X_1^3\nonumber\\
&& +\Biggl(\frac{103}{2520} \ln(2\omega r \eta) -\frac{824923}{6350400}+\frac{1}{8} \gamma^2
-\frac{7}{432}\pi^2-\frac{43}{432}i\pi-\frac{883}{5040}\gamma+\frac{1}{12}i\pi\gamma \Biggl) X_2^4 X_1^2\nonumber\\
&&+\left(-\frac{1}{5670} \ln(2\omega r \eta) +\frac{170659}{228614400}-\frac{1}{2268}\gamma \right) 
X_2^5X_1+\frac{1}{14515200}X_2^6
\,.
\end{eqnarray}
\end{widetext}
Note that this ``up'' solution contains logarithmic terms,  $\ln(2\omega r \eta)$, 
starting at the 3PN level  (i.e.  $A_{6}^{\rm up}$),  and even squared logarithms, $(\ln(2\omega r \eta))^2$, at the 6PN level $A_{12}^{\rm up}$.  These logarithms have various physical meanings: some are gauge effects, some are related to
far-zone effects and enter the relation between algorithmic multipole moments and source variables\footnote{This is the
case of the $\ln \eta$ appearing at 3PN; see \cite{Anderson:1982fk,Blanchet:1987wq}.}, some (with imaginary coefficients) are linked to tail modifications of radiation-reaction effects, and some are linked to tail modifications of
conservative effects. For instance, the term $-\frac{4}{45}\ln(2\omega r \eta) X_2^3X_1$, which contains a factor
$M \omega^6$, is encoding the conservative part of the 4PN near-zone hereditary term discussed in 
 \cite{Blanchet:1987wq,Damour:2009sm,Blanchet:2010zd}. Our computation of the 
 quantity $h_{kk}$ will automatically select, among all these logarithmic contributions, the  gauge-invariant
 conservative ones.

 As for the ingoing hypergeometric solution $X_{l\omega}^{({\rm in})} (r)$ we found that, modulo an inessential constant prefactor, it is correctly evaluated by solving the corresponding homogeneous RW equation by a formal PN scheme,
 of the type we explain next.
 
 \subsection{Deriving sufficiently accurate homogeneous PN-expanded solutions   for general multipole degree $l$  }
 
The hypergeometric solutions   (\ref{eq36}), (\ref{eq37}) automatically incorporate all the physically correct boundary
conditions, but are difficult to evaluate for a general multipole degree $l$. By contrast, for generic $l$,
one can look for solutions of the homogeneous RW equations 
${\mathcal L}_{({\rm RW})}^{(r)} X_{({\rm in})} (r) = 0 = {\mathcal L}_{({\rm RW})}^{(r)} X_{({\rm up})} (r)$
 satisfied by
both  $X_{l\omega}^{({\rm in})}$ and $X_{l\omega}^{({\rm up})}$ in the form of a usual PN expansion, i.e.
\begin{eqnarray}
\label{eq41}
X_{l\omega}^{{\rm in} ({\rm PN})} (r) &= &r^{l+1} [1+\eta^2 A_2^{(PN, l)} + \eta^4 A_4^{(PN, l)}  \\
&+&\eta^6 A_6^{(PN, l)} + \ldots + \eta^{12} A_{12}^{(PN, l)} + \ldots] \, ,\nonumber
\end{eqnarray}
\begin{eqnarray}
\label{eq42}
X_{l\omega}^{{\rm up} ({\rm PN})} (r) &= &r^{-l} [1+\eta^2 A_2^{(PN,-l-1)} + \eta^4 A_4^{(PN,-l-1)}  \\
&+&\eta^6 A_6^{(PN, -l-1)} + \ldots + \eta^{12} A_{12}^{(PN, -l-1)} + \ldots] \, ,\nonumber
\end{eqnarray}
As the coefficients entering the RW operator  ${\mathcal L}_{({\rm RW})}^{(r)} $  (\ref{eq24}) are polynomial in $GM/c^2=M\eta^2$ and 
$\omega^2/c^2=\omega^2 \eta^2$, one can look for PN expansion coefficients 
$A_2^{(PN, l)} , A_4^{(PN, l)} , \ldots$ (restricted, as indicated in Eqs.~(\ref{eq41}),(\ref{eq42}), to even powers of $\eta=1/c$, i.e. to even values of the index $k$) in the form of  polynomials in $X_1 = GM/r$, $X_2 = (\omega r)^2$ with $l$-dependent coefficients. One finds that it is indeed possible to find such solutions. The first three coefficients of such PN-expanded
solutions read
\begin{eqnarray}
A_0^{{\rm in}\, (PN,\, l)}&=& 1\nonumber\\
A_2^{{\rm in}\, (PN,\, l)}&=& -\frac{(l-2)(l+2)}{l}X_1 -\frac{1}{2(2l+3)} X_2\nonumber\\
A_4^{{\rm in}\, (PN,\, l)}&=&\frac{(l-2)(l-3)(l+2)(l+1)}{(-1+2l)l} X_1^2 \nonumber\\
&&+\frac{(l^3-5l^2-14l-12)}{2(l+1)(2l+3)l}X_2X_1 \nonumber\\
&&+\frac{1}{8(2l+5)(2l+3)}X_2^2\nonumber\\
A_6^{{\rm in}\, (PN,\, l)}&=& -\frac{(l-2)(l-3)(l-4)(l+2)(l+1)}{3 (-1+2l)(l-1)}  X_1^3\nonumber\\
&&-\frac{2(15 l^4+30 l^3+28 l^2+13 l+24)}{(-1+2l) (2l+1)(l+1)l(2l+3)} \times\nonumber\\
&&\times X_1^2 X_2 \ln(r/R) \nonumber\\
&&-\frac{ (3 l^4-27 l^3-154 l^2-220 l-120)}{24 (l+1) l (2 l+5) (l+2) (2 l+3)} X_1 X_2^2\nonumber\\
&&-\frac{1}{48(2 l+5) (2 l+7) (2 l+3)} X_2^3 
\end{eqnarray}
As illustrated here for the coefficient  $A_6^{(PN, l)}$, it is necessary to go beyond a purely polynomial structure
in   $X_1 = GM/r$ and  $X_2 = (\omega r)^2$, and to introduce, in a few terms,
a logarithmic dependence $A_{2k, p,q}^{(PN, l) \log} \ln (r/R) X_1^p X_2^q$ in 
the coefficients $A_{2k}^{(PN, l)}$ entering  the possible form of the PN expansion of  $X_{l\omega}^{\rm in} $. 
Such logarithmic terms must be included in $A_6^{(PN, l)}$ , $A_8^{(PN, l)}$,  $A_{10}^{(PN, l)}$ , and $A_{12}^{(PN, l)}$ . In addition, as indicated in  Eq.~(\ref{eq42}) above, the coefficients entering the {\it up} PN solution $X_{l\omega}^{{\rm up} ({\rm PN})}$, Eq.~(\ref{eq42}), are formally obtained from the $A_{2k}^{(l)}$ coefficients entering the {\it in} PN solution Eq.~(\ref{eq41}) simply by replacing $l$ by $-l-1$. However, this formal replacement rule sometimes generates  poles 
$\propto (l-n)^{-1}$ where $n$ is a positive integer. Such poles correspond to a denominator of the form
$(l+n+1)^{-1}$ in some coefficient  $A_{2k}^{(PN, l)}$ in $ X_{l\omega}^{{\rm in} ({\rm PN})} $.  For instance, the 4PN
``in'' coefficient  $A_{8}^{(PN, l)}$ contains the following term 
\begin{eqnarray}
+\frac{ (5 l^5-60 l^4-645 l^3-1788 l^2-1928 l-840)}{240 l (l+3) (2 l+3) (l+1) (2 l+5) (2 l+7) (l+2)} X_1 X_2^3\nonumber\\
\end{eqnarray}
This term (which is $\propto X_1 X_2^3$) contains a factor  $l+3$ in the denominator that will generate
a  denominator $l-2$ in the corresponding 4PN-level term $A_{8}^{(PN,- l - 1)}$  of  $X_{l\omega}^{{\rm up} ({\rm PN})} $.
When considering the integer value $l=2$, one must replace the ill-defined term $\propto X_1 X_2^3/(l-2)$ in
$X_{l\omega}^{{\rm up} ({\rm PN})} $ by a new logarithmic  term   $\propto X_1 X_2^3   \ln (r/R)  $.
Similar additional logarithms in the ``up'' solution  also arise (for the multipole $l=2$) from the ``in'' coefficients   $A_{10}^{(PN, l)}$  and  $A_{12}^{(PN, l)}$.  Analogous additional logarithms also arise for   $l=3$ from  $A_{10}^{(PN, l)}$  and  $A_{12}^{(PN, l)}$, and for $l=4$ from   $A_{12}^{(PN, l)}$.  

The so generated PN solutions depend on the arbitrary scale $R$ (entering the 
two sorts of logarithms discussed above). Moreover, these formal solutions do not correctly incorporate the 
retarded-type boundary condition included in the hypergeometric solutions   (\ref{eq36}), (\ref{eq37}). [For instance,
the contributions proportional to odd powers of $\eta$ in (\ref{eq36}) come from the retarded nature of the ``up'' solution.]
However, we have checked (by comparing the structure of the hypergeometric expansions to that of
the simpler PN expansions) that,  for $l \ge 5$, they give the same value for $h_{kk}$ (up to 6PN).  
[In particular,  the arbitrary logarithmic   scale $R$ drops out in the 6PN calculation
of $h_{kk}$.]
Moreover, for
$l=2$,  $l=3$ and $l=4$, where one must use the hypergeometric-type solution for   $X_{l\omega}^{({\rm up})} (r)$,
it would be enough to use the PN-one for  $X_{l\omega}^{({\rm in})} (r)$. 
Indeed, an explicit calculation shows that  they
are proportional to each other (up to $\eta^{12}$ included).  For instance, we found
\be
X_{\rm (in)}^{HG, l=2}=\frac{ie^{\psi_2}}{384M^4\omega}X_{\rm (in)}^{PN, l=2}\
\ee
where 
\begin{eqnarray}
\psi_2 &=& -\frac{94}{35}iM\omega\eta^3-\left[\frac{24197}{4900}+\frac{214}{105}\ln\left(\frac{R}{2M\eta^2}\right)\right]\omega^2 M^2\eta^6 \nonumber\\
&& -\left(\frac{1099309}{385875}+\frac{428}{315} \pi^2\right) i M^3\omega^3\eta^9\nonumber\\
&&+\left[-\frac{832550660525191}{44532259380000} +\frac{1712}{105} \zeta(3)\right.\nonumber\\
&& \left. +\frac{45796}{33075} \pi^2  -\frac{3390466}{1157625} \ln\left(\frac{R}{2M\eta^2}\right)\right]M^4\omega^4\eta^{12}\,
\end{eqnarray}
and $R$ is the arbitrary (constant) scale entering the logarithmic terms in the PN solution. The 6PN-accurate identity (modulo an overall factor) between  $X_{\rm (in)}^{HG, l}$  and   $X_{\rm (in)}^{PN, l}$ follows from the result
of  \cite{Poisson:1994yf,Sasaki:2003xr} that the horizon boundary condition does not influence the PN-expanded, 
near-zone metric until terms of order $O(\eta^{13})$.

\subsection{Summing over $m$}

When inserting the results (\ref{eq39}), (\ref{eq40}), (\ref{eq41}), (\ref{eq42}) in the expressions (of the type (\ref{eq35odd}),
(\ref{eq35even})) giving $h_{kk,lm}$ we get explicit results which depend both on $l$ and on $m$ via the $m$-dependent value of $\omega = m\Omega$. Indeed, $\omega$ appeared from the start in the RW equations, and thereby in their solutions; e.g., note that the parameter $\epsilon = 2GM\omega/c^3$ pervasively entering the hypergeometric expansions (\ref{eq36}), (\ref{eq37}) (and the $l$-deformed parameter $\nu$, Eq.~(\ref{eq38})) is proportional to $m$. The summation over $m$ in Eq.~(\ref{eq22}) then generates finite sums most of which are of the form
\be
\label{eq43}
S_{N,l} = \sum_{m=-l}^{+l} m^N \vert Y_{lm} (\pi/2,0)\vert^2 \, ,
\ee
or
\be
\label{eq44}
S'_{N,l} = \sum_{m=-l}^{+l} m^N \vert \partial_{\theta} Y_{lm} (\pi/2,0)\vert^2 \, .
\ee
These sums vanish when the (non negative) integer $N$ is odd, and can be expressed as polynomials in $l$ when $N$ is even, thanks to the results of the Japanese relativity school ; see Appendix F in \cite{Nakano:2003he}.
Actually, we needed to go beyond the explicit results given in the latter reference: indeed, we needed $S_{N,l}$
up to $N=12$, and  $S'_{N,l}$ up to $N=10$. The explicit values of the above sums for $l=2$ and 3 are:
\begin{eqnarray}
S_{N,2}&=& \sum_{m=-2}^2 m^N \vert Y_{2m} (\pi/2,0)\vert^2\nonumber\\
&=& \frac{15}{32\pi} [(-2)^N+2^N]\nonumber\\
S_{N,3}&=& \sum_{m=-3}^3 m^N \vert Y_{3m} (\pi/2,0)\vert^2\nonumber\\
&=& \frac{35}{64\pi}[(-3)^N+3^N]+\frac{21}{64\pi}[(-1)^N+1]\,,
\end{eqnarray}
and
\begin{eqnarray}
S'_{N,2} &=& \sum_{m=-2}^{+2} m^N \vert \partial_{\theta} Y_{2m} (\pi/2,0)\vert^2\nonumber\\
&=& \frac{15}{8\pi}[(-1)^N+1] \nonumber\\
S'_{N,3} &=& \sum_{m=-3}^{+3} m^N \vert \partial_{\theta} Y_{3m} (\pi/2,0)\vert^2\nonumber\\
&=& \frac{105}{32\pi}[(-2)^N+2^N]\,.
\end{eqnarray}

In addition to the sums (\ref{eq43}), (\ref{eq44}), our 6PN-accurate calculation of $h_{kk}^R$ involved  new, and more delicate, sums of the type
\be
\label{eq45}
S^{\log}_{N,l} = \sum_{m=-l}^{+l} m^N \ln (-im) \vert Y_{lm} (\pi/2,0)\vert^2  \, .
\ee
Our 6PN calculation involve such logarithmic sums for the $l$ values $l=2$, $l=3$ and $l=4$. 
For $l=2$ and $l=3$, we have for example
\begin{eqnarray}
S^{\log}_{N,2} &=& \sum_{m=-2}^{+2} m^N \ln (-im) \vert Y_{2 m} (\pi/2,0)\vert^2\nonumber\\
&=& \frac{15}{32\pi}(2^N+(-2)^{N})\ln 2 -\frac{15}{64}i (2^N-(-2)^{N})\,\nonumber\\
S^{\log}_{N,3} &=& \sum_{m=-3}^{+3} m^N \ln (-im) \vert Y_{3 m} (\pi/2,0)\vert^2\nonumber\\
&=& \frac{35}{64\pi}(3^N+(-3)^{N})\ln 3 \nonumber\\
&&-\frac{35}{128}i (3^N-(-3)^{N}) -\frac{21}{128}i (1-(-1)^N)\,.
\end{eqnarray}
Contrary to the non-logarithmic sums above, these logarithmic sums do not vanish for odd values of $N$.
Separating the cases where $N$ is even or odd we  have  
\begin{eqnarray}
S^{\log}_{2N,2} 
&=& \frac{15}{16\pi} 2^{2N}\ln 2 \,,\quad
S^{\log}_{2N+1,2} 
= -\frac{15}{32} i 2^{2N+1}\,.
\end{eqnarray}
and
\begin{eqnarray}
S^{\log}_{2N,3} 
&=& \frac{35}{32\pi} 3^{2N}\ln 3 \,,\quad
\nonumber\\
S^{\log}_{2N+1,3} 
&=& -\frac{7i}{64}\left(5\cdot  3^{2N+1}+3  \right)\,.
\end{eqnarray}
These sums explain the appearance of  $\ln 2$ and $\ln 3$ in our final results.
Note also that the squared logarithms that entered the 6PN ``up'' coefficients $A_{12}^{{\rm up}\, (HG,\, l)}$
do not lead (at the 6PN order) to a conservative contribution involving a sum of the type (\ref{eq45}) with
$ \ln (-im) $ replaced by its square.

\subsection{Subtracting   $\tilde D_0$, and summing the series over the multipolar degree $l$}

After having explicitly performed the summation over $m$, we need to subtract from each $l$-contribution the $u$-expansion of $\tilde D_0 = (1-3u) D_0$ (with Eq.~(\ref{eq21})), namely
\begin{eqnarray}
\tilde D_0&=  &2u-\frac{13}{2}u^2+\frac{9}{32}u^3+\frac{83}{128}u^4+\frac{12361}{8192}u^5\nonumber\\
&+&\frac{116163}{32768}u^6 +\frac{1867635}{32768}u^7 + O(u^8) \,.
\end{eqnarray}
We then obtain, according to Eq.~(\ref{eq22}), an explicit expression for $h_{kk}^R$ given by a sum of 
the following type: (i) a few exact contributions for $l=0$ and $l=1$; (ii) the explicit contributions from $l=2$, $l=3$ 
and $l=4$
(expanded up to $\eta^{12}$); and, finally, (iii) an infinite series over $l \geq 4$. The latter series converges, and can be 
explicitly evaluated thanks to the fact that the PN-expanded RW solutions for generic values of $l$ can be written in terms of 
explicit {\it rational functions} of $l$. In other words, our calculation involves series of the type 
$\sum_{l \geq 3} P_{n-2} (l)/Q_n(l)$ with (complicated) polynomials of degree $n-2$ and $n$ respectively. For instance, at 
the 4PN level we have a polynomial $Q_{14} (l)$ of degree 14 in the denominator. [However, $Q_{14} (l)$ is factorizable in
 ten simpler factors $\sim (al+b)^k$ with $k = 1$ or $2$ and $a, b \in {\mathbb Z}$.] In the course of this calculation, one checks 
 that $\tilde D_0$ does precisely subtract the $l \to \infty$ piece in the corresponding {\it unsubtracted} form of $h_{kk}^{(l)}$,
  which, e.g., involves at the 4PN level a rational fraction of the type $P_{14} (l)/Q_{14} (l)$.

The convergent series entering our calculation can be evaluated (after decomposing it in partial fractions in $l$) in terms of 
the well-known Euler series $\zeta (2) = \sum_l 1/(l+1)^2 = \pi^2 / 6$, and $\zeta(4)= \sum_l 1/(l+1)^4 = \pi^4 /90$.

This finally leads  to our 6PN-accurate result
 \begin{eqnarray}
 \label{hkk6PN}
h_{kk}^{R}&=&-2u+5u^2+\frac{5}{4}u^3+\left(-\frac{1261}{24}+\frac{41}{16}\pi^2\right)u^4\nonumber\\
& +&\Biggl(\frac{157859}{960}-\frac{256}{5}\gamma-\frac{128}{5}\ln(u)-\frac{512}{5}\ln(2)\nonumber\\
&&-\frac{2275}{256}\pi^2\Biggl)u^5\nonumber\\
&+&\Biggl(\frac{284664301}{201600}+\frac{28016}{105}\gamma+\frac{14008}{105}\ln(u)\nonumber\\
&&+\frac{63472}{105}\ln(2)-\frac{246367}{1536}\pi^2-\frac{486}{7}\ln(3)\Biggl) u^6\nonumber\\
&& {\pmb -}\frac{27392}{525}\pi u^{13/2} \nonumber\\
& +&\Biggl(-\frac{413480}{567}\ln(2)+\frac{5044}{405}\ln(u)+\frac{10088}{405}\gamma\nonumber\\
&&+\frac{22848244687}{7257600}+\frac{4617}{7}\ln(3)\nonumber\\
&& +\frac{2800873}{131072}\pi^4-\frac{608698367}{884736}\pi^2\Biggl)u^7 + o(u^7)
\,.
\end{eqnarray}

\section{6PN-accurate computation of the EOB radial potential $A(u;\nu)$ at linear order in $\nu$}
\label{sec:three}

Let us now translate our results on Deweiler's gauge-invariant function $h_{kk}^R(u)$ into the crucial
radial interaction potential $A(r;m_1,m_2)$ entering the EOB framework.
 The potential $A(r;m_1,m_2)$ is a {\it gauge-invariant} function which enters the EOB formalism \cite{Buonanno:1998gg,Buonanno:2000ef,Damour:2000we,Damour:2001tu}. It is a useful generalization of the well-known Schwarzschild potential $A^S (r) = 1-2GM/c^2r$. The EOB formalism maps the {\it conservative} dynamics of a (non spinning) two-body system $(m_1,m_2)$ onto the geodesic dynamics of one body of mass $\mu = m_1m_2 / (m_1 + m_2)$ in a stationary and spherically symmetric ``effective'' metric,
\begin{eqnarray}
\label{eq1}
ds_{\rm eff}^2 &= &-A (r;m_1,m_2) c^2 dt^2 \nonumber \\
&+ &B(r;m_1,m_2) dr^2 + r^2 (d\theta^2 + \sin^2 \theta d\varphi^2) \, ,
\end{eqnarray}
together with post-geodesic corrections described by a function $Q(r,p_r , p_{\varphi} ; m_1 , m_2)$ which is, at least, quartic in the (EOB) radial momentum $p_r$ \cite{Damour:2000we}. The latter (gauge-fixing) restriction ensures that the gauge-invariant dynamics of the sequence of circular orbits is fully encoded in the sole radial potential $A(r;m_1,m_2)$ \cite{Damour:2009sm,Barausse:2011dq,Akcay:2012ea}.
 One can write the  radial potential $A(r;m_1,m_2)$
as a function of the two variables  $u := GM/c^2r$  and $\nu= \mu/M$:  $A(r;m_1,m_2) \equiv A(u;\nu)$.

From the analytical point of view, one can  approach this function
of two variables in two different, and complementary, ways: either by expanding it in powers of $u$, or in powers of $\nu$.
The expansion in $u$ corresponds to the usual PN expansion, while the expansion in $\nu$ is a GSF expansion. 
In addition, one can also approach the radial potential  $A(u;\nu)$ by several types of numerical simulations: either
full, three-dimensional numerical relativity simulations, or GSF numerical simulations.
In this paper, we shall focus on analytical knowledge, without dealing with its comparison with numerical results.

Before discussing the new, higher PN contributions to $A(u;\nu)$ that we shall derive below,
let us recall the present PN knowledge of $A(u;\nu)$. Its PN expansion is completely known only up to  
the 4PN level (included), i.e. up to terms of order $u^5$. 
At the 5PN level (i.e. $O(u^6)$), only the logarithmic contributions are so far known.
At this order, the radial potential has the form
\begin{eqnarray}
\label{eq2}
A(u;\nu) &= &1-2u + \nu a_3(\nu) u^3 + \nu a_4 (\nu) u^4 \nonumber \\
&+&  \nu (a_5^c (\nu) + a_5^{\ln} (\nu) \ln u) u^5 \nonumber\\
&+&  \nu (a_6^c (\nu) + a_6^{\ln} (\nu) \ln u) u^6 +  o(u^6) \, .
\end{eqnarray}
There is no 1PN-level contribution (i.e. $a_2 (\nu) = 0$). The value of the 2PN-level coefficient, namely 
\be
\label{eq3}
a_3 (\nu) = 2 \, ,
\ee
was derived in \cite{Buonanno:1998gg} from the 2PN Delaunay Hamiltonian of \cite{Damour:1988mr}. The value of the 3PN-level coefficient, namely
\be
\label{eq4}
a_4 (\nu) = \frac{94}3 - \frac{41}{32} \, \pi^2 \, ,
\ee
was derived in \cite{Damour:2000we} from the 3PN Hamiltonian of \cite{Jaranowski:1997ky,Damour:2001bu}. The value of the 4PN-level {\it logarithmic} coefficient, namely
\be
\label{eq5}
a_5^{\ln} (\nu) = \frac{64}5 \, ,
\ee
was derived in \cite{Damour:2009sm,Blanchet:2010zd,Barausse:2011dq} from the results of Ref.~\cite{Blanchet:1987wq}.
The value of the non-logarithmic 4PN-level coefficient in Eq.~(\ref{eq2}) was derived in \cite{Bini:2013zaa},
and found to be equal to (with $\gamma$ denoting Euler's constant)
\begin{subequations}  \label{eqs6}
\begin{eqnarray}
\label{eq6a}
a_5^c (\nu) &= &a_5^{c0} + \nu a_5^{c1}  \, ,\\
\label{eq6b}
a_5^{c0} &=  &-\frac{4237}{60} + \frac{2275}{512} \pi^2 + \frac{256}5 \ln 2 + \frac{128}5 \gamma \, ,\\
\label{eq6c}
a_5^{c1} &= &-\frac{221}6 + \frac{41}{32} \pi^2 \, .
\end{eqnarray}
\end{subequations}

Finally, at the 5PN level, only the logarithmic contribution $ \nu  a_6^{\ln} (\nu) \ln u \,  u^6 $ has been obtained
 \cite{Blanchet:2010zd,Damourlogs,Barausse:2011dq}, and found to be given by
\be
 a_6^{\ln} (\nu) = -\frac{7004}{105} -\frac{144}{5} \nu
\ee

In the present paper, we shall extend our analytical knowledge of  $A(u;\nu)$ to the 6PN level, but only for the
terms linear in $\nu$.  Our results are then most simply expressed in terms of  the first coefficient function $a(u)$ of 
the GSF expansion of the radial potential $ A(u;\nu)$:
\be
\label{eq2a}
A(u;\nu) = 1-2u + \nu a(u) + \nu^2 a_2 (u) +   \nu^3 a_3(u) + O(\nu^4) \, .
\ee
 The  terms  $\nu a(u) + \nu^2 a_2 (u) + \cdots$ represent
 (from the point of view of GSF theory)  corrections  to the test-mass limit 
 (i.e. $A^S (u) = 1-2GM/c^2r =1- 2 u$) coming from GSF effects.
 Note that, using the link (\ref{avshkk}),  Refs. \cite{Barausse:2011dq} and \cite{Akcay:2012ea} gave numerical estimates of the EOB function $a(u)$ beyond the weak-field (PN) regime $u \ll 1$. In particular, Akcay et al. \cite{Akcay:2012ea} gave accurate numerical representations of the function $a(u)$ over the interval $0 < u < 1/3$, and discovered the presence of a singularity near the ``light-ring'' $u \to 1/3$.

 Previous work has shown that remarkable cancellations occur in the EOB
 potential  $A(u;\nu)$: while related functions (such as the energy-versus-frequency function $E(x; \nu)$) generally contain,
 at each PN order, all the a priori possible powers of $\nu$, the PN expansion of   $A(u;\nu)$ is linear in $\nu$ up to
 the 3PN order (included) \cite{Buonanno:1998gg,Damour:2000we}; in other words, the $O(\nu^2)$ term $a_2 (u)$ in Eq.~(\ref{eq2a}) starts only at the 4PN order $\sim u^5$ \cite{Bini:2013zaa}. The latter work also showed the presence of
 cancellations at the 4PN order; namely, the $u^5$ contribution to  $A(u;\nu)$ is no more than quadratic in $\nu$
 (by contrast with, e.g., the energy-frequency function $E(GM\Omega/c^3;\nu)$  which has $O(\nu^3)$ and $O(\nu^4)$ contributions \cite{Jaranowski:2012eb}). In other
 words, the  $O(\nu^3)$ term $a_3 (u)$ in Eq.~(\ref{eq2a}) does not start before the 5PN order $u^6$.]
 We shall further discuss these cancellations below.
 
Inserting our result Eq. (\ref{hkk6PN}) for $h_{kk}^R$ in  the link (\ref{avshkk}), we determined the PN expansion of the 1GSF coefficient $a(u)$ up to the 6PN level, i.e.
 \begin{eqnarray}
 \label{eq2c}
 a(u)&= &a_3 u^3 + a_4 u^4 + a_5(\ln u) u^5+ a_6(\ln u) u^6 \nonumber\\ && 
 +a_{6.5} u^{13/2} +a_7(\ln u) u^7 + o(u^7)
 \end{eqnarray}
The coefficients $a_3, \cdots, a_6(\ln u)$ here correspond to the $\nu \to 0$ limits of the PN-expansion coefficients
entering  Eq.~(\ref{eq2}) above; i.e. $a_3=a_3(0), a_4=a_4(0), a_5(\ln u) = a_5^{c}(0) +  a_5^{\ln} (0) \ln u,
a_6(\ln u) = a_6^{c}(0) +  a_6^{\ln} (0) \ln u$.  For these coefficients, our results confirm the previous determinations
of  $a_3(0), a_4(0),  a_5^{c}(0) , a_5^{\ln} (0)$ and $ a_6^{\ln} (0) $ recalled above, and extend them by providing
the analytical values of  $a_6^{c}(0)$,  $a_{6.5}$ and  $a_7(\ln u) = a_7^{c}(0) +  a_7^{\ln} (0) \ln u$. Namely, we found
\begin{subequations}  \label{eqs6pn}
\begin{eqnarray}
\label{eq6pna}
a_6(\ln u) &= &-\frac{1066621}{1575}-\frac{14008}{105}\gamma-\frac{7004}{105}\ln(u)\nonumber\\
&&-\frac{31736}{105}\ln(2)+\frac{246367}{3072}\pi^2+\frac{243}{7}\ln(3) \\
\label{eq6pnb}
a_{6.5} &=  &+\frac{13696}{525}\pi  \, ,\\
\label{eq6pnc}
a_7(\ln u)&= & \frac{206740}{567}\ln(2)-\frac{2522}{405}\ln(u)-\frac{5044}{405}\gamma\nonumber\\
&&-\frac{1360201207}{907200}-\frac{4617}{14}\ln(3)  
-\frac{2800873}{262144}\pi^4\nonumber\\
&&+\frac{608698367}{1769472}\pi^2 \, .
\end{eqnarray}
\end{subequations}
Note that the transcendentality of the coefficients $a_n$ increases with $n$: $a_3$ is rational; $a_4$ involves $\zeta(2)=\pi^2/6$; $a_5$ involves   $\zeta(2)$, $\ln 2$ and $\gamma$;  $a_6$ involves   $\zeta(2)$, $\ln 2$,  $\gamma$
and $\ln 3$; while  $a_7$ involves   $\zeta(2)$, $\ln 2$,  $\ln 3$, $\gamma$, and $\zeta(4)$. Formally,  writing the
Fokker action of the binary system in terms
of classical Feynman-like diagrams \cite{Damour:1995kt} leads to integral expressions for the expansion coefficients of
the radial potential $A(u, \nu)$ which are of the type called ``periods'' in mathematics. However, for circular orbits, these periods 
involve (non algebraic) helical world lines (which explains the appearance of Euler's constant $\gamma$).

\subsection{Binding energy of binary systems in terms of the orbital frequency}

The (main) EOB radial potential $A(u;\nu)$ (which defines the $g_{00} (r)$ component of the effective metric entering the EOB formalism) completely describes the gauge-invariant dynamics of circular orbits of binary systems. 
In particular, the potential $A(u;\nu)$ allows one to give a simple parametric representation of the functional link between the total energy $H^{\rm tot}$ of a binary system and the frequency of a circular orbit \cite{Buonanno:2000ef,Damour:2009sm,Barausse:2011dq, Akcay:2012ea}. Let us briefly recall this parametric representation, and  use it to determine
the functional relation between the energy and the orbital frequency (or, more conveniently, the dimensionless frequency
parameter $x = (M\Omega)^{2/3}$).

Given the EOB potential $A(u;\nu)$ , the total energy $H^{\rm tot}$ can be computed as an explicit function of $u$. Let us first define the functions
\be
\label{eq50}
\tilde A (u;\nu) := A(u;\nu) + \frac12 \, u \, \partial_u \, A(u;\nu) \, ,
\ee
\be
\label{eq51}
\hat H_{\rm eff} (u;\nu) := \frac{A(u;\nu)}{\sqrt{\tilde A (u;\nu)}} \, ,
\ee
\be
\label{eq52}
h (u;\nu) := \sqrt{1 + 2 \nu (\hat H_{\rm eff} (u;\nu) - 1)} \, .
\ee
In terms of this notation, the total energy of the circular binary (with $c=1$) is simply $H^{\rm tot} (u;\nu) = Mh(u;\nu)$ so that the binding energy $E_B = H^{\rm tot} - M$ reads
\begin{eqnarray}
\label{eq53}
E_B (u;\nu) &= &M (h(u;\nu) - 1) \nonumber \\
&= &M \left( \sqrt{1+ 2\nu (\hat H_{\rm eff} (u;\nu) - 1)} -1 \right) \, .
\end{eqnarray}
On the other hand, the dimensionless frequency parameter $x = (M\Omega)^{2/3}$ is given by the following function of $u$
\be
\label{eq54}
x(u;\nu) = u \left( \frac{-\frac12 \, \partial_u \, A (u;\nu)}{h^2 (u;\nu)} \right)^{1/3} \, .
\ee

Up to now we have made no approximation, so that Eqs.~(\ref{eq53}) and (\ref{eq54}) yield an {\it exact} parametric representation of the functional link between $ E_B$ and $x$. If we now consider the PN expansion of $A (u;\nu)$ up to some PN level, we get corresponding PN expansions of $ E_B(u)$ and $x(u)$. Inverting the latter expansion (which starts as $x(u) =  u + \frac13 \, \nu u^2 + O(u^3)$) to get $u$ in terms of $x$, we can straightforwardly obtain the PN expansion of the function relating $ E_B$ to the frequency parameter $x$. 

Let us start from the PN expansion of  $A(u;\nu)$ written up to the 6PN level, i.e.
\begin{eqnarray}
&&A(u; \nu)=1-2 u+\nu a_3(\nu) u^3+\nu a_4(\nu) u^4+\nu (a_{5}^c(\nu)\nonumber\\
&&\qquad +a_5^{\ln}(\nu)\ln(u)) u^5
+\nu (a_6^c(\nu)+a_6^{\ln} (\nu)\ln(u)) u^6\nonumber\\
&&\qquad +\nu a_{6.5}(\nu) u^{13/2}+\nu (a_{7}^c(\nu)+a_7^{\ln}(\nu)\ln(u)) u^7
\end{eqnarray}
Here, we assumed that there are no extra terms that have not been detected by our $\nu$-linear computation
at 5, 5.5 and 6PN; for instance, we assume that there are no terms of the type $\nu^2 ( \ln u)^2$  in the higher PN
coefficients  $a_n(\nu, \ln u)$. 
Let us use in this expansion all the information we currently have; i.e., let us replace the  known coefficients 
 $a_n(\nu, \ln u)$ (up to $n=5$), as well as $a_{6}^{\ln}(\nu)$, by their known values 
(recalled at the beginning of this Section), and let us replace the coefficients which are only known
 to linear order in $\nu$ by general expressions of the type:
$a_{6}^c(\nu) =a_{6}^c(0)+\nu a_{6}^{c1}(\nu)$,
$ a_{6.5}(\nu)= a_{6.5}^c(0)+ \nu a_{6.5}^{c1}(\nu) $,
$a_{7}^c(\nu) =a_{7}^c(0)+\nu a_{7}^{c1}(\nu)$,
$a_{7}^{\ln}(\nu) =a_{7}^{\ln}(0)+\nu a_{7}^{\ln 1}(\nu)$,\\
where all the coefficients  $a_{6}^c(0)$,  $a_{6.5}^c(0)$,  $a_{7}^c(0)$,  $a_{7}^{\ln}(0)$ have been
determined above, and will be henceforth replaced by their explicit values.
Note that we have allowed here for an a priori arbitrary $\nu$ dependence of the terms of order $O(\nu^2)$
that are not controlled by our present results.  Actually, we expect that the $\nu$ dependence of these coefficients
will be rather simple. See below.

With such a parametrization of our current knowledge (and of our remaining ignorance), we find that the coefficients
in the 6PN expansion  Eq.~(\ref{eq10}) of  the function $ E_B(x;\nu) $  are given by the following expressions
(in which we suppress, for notational clarity, the arguments $\nu$ in  $a_{6}^{c1}(\nu)$,   $  a_{6.5}^{c1}(\nu)$, $a_{7}^{c1}(\nu)$,  and $ a_{7}^{\ln 1}(\nu)$, and separate the logarithmic contributions as
$e_{\rm nPN}(\nu, \ln x) \equiv e_{\rm nPN}^c(\nu) + e_{\rm nPN}^{\ln}(\nu)  \ln x$) 
\begin{widetext} 
\begin{eqnarray}
\label{eB6PN}
e_{\rm 1PN} (\nu)&=&  -\frac{1}{12}\nu-\frac34 \nonumber\\
e_{\rm 2PN} (\nu)&=&  -\frac{1}{24}\nu^2+\frac{19}8 \nu-\frac{27}{8}\nonumber\\
e_{\rm 3PN} (\nu)&=& -\frac{35}{5184}\nu^3 -\frac{155}{96}\nu^2+\frac{34445}{576}\nu-\frac{205}{96}\pi^2\nu-\frac{675}{64}  \nonumber\\
e_{\rm 4PN}^c (\nu)&=& -\frac{3969}{128}+\frac{77}{31104}\nu^4+\frac{301}{1728}\nu^3
+\left(-\frac{498449}{3456} +\frac{3157}{576}\pi^2\right)\nu^2\nonumber\\
&+&\Biggl(-\frac{123671}{5760}+\frac{1792}{15} \ln(2)+\frac{9037}{1536} \pi^2
+\frac{896}{15} \gamma\Biggl)\nu \nonumber\\
e_{\rm 4PN}^{\ln}(\nu)&=& \frac{448}{15}\nu  \nonumber\\
e_{\rm 5PN}^c (\nu)&=& -\frac{45927}{512}+\frac{1}{512}\nu^5+\frac{55}{512}\nu^4
+\left(-\frac{1353}{256}\pi^2+\frac{69423}{512} \right)\nu^3\nonumber\\
&+&\left(-\frac{21337}{1024}\pi^2+3 a_6^{c1}
-\frac{896}{5}\ln(2)-\frac{448}{5}\gamma+ \frac{893429}{2880} \right)\nu^2\nonumber\\
&+&\left(-\frac{228916843}{115200}-\frac{9976}{35}\gamma+\frac{729}{7}\ln(3) 
 -\frac{23672}{35}\ln(2)+\frac{126779}{512}\pi^2 \right)\nu \nonumber\\
e_{\rm 5PN}^{\ln}(\nu)&=& -\frac{4988}{35}\nu- \frac{656}{5} \nu^2\nonumber\\
e_{\rm 5.5PN}(\nu)&=&\frac{10}{3}\nu \left(\frac{13696}{525}\pi+\nu  a_{6.5}^{c1}\right)\nonumber\\
e_{\rm 6PN}^c (\nu)&=& -\frac{264627}{1024}+\frac{2717}{6718464}\nu^6+\frac{5159}{248832}\nu^5
+\left(\frac{272855}{124416}\pi^2-\frac{20543435}{373248} \right)\nu^4\nonumber\\
&+&\Biggl(\frac{1232}{27}\gamma+\frac{6634243}{110592}\pi^2-\frac{11}{2} a_6^{c1}
- \frac{71700787}{51840} +\frac{2464}{27}\ln(2) \Biggl)\nu^3\nonumber\\
&+& \Biggl( \frac{113176680983}{14515200} +\frac{18491}{2304}\pi^4
+\frac{246004}{105}\ln(2)+\frac{112772}{105}\gamma+\frac{11}{2} a_6^{c1}+\frac{2}{3}a_7^{\ln 1}\nonumber\\
&+& \frac{11}{3} a_7^{c1}-\frac{86017789}{110592}\pi^2-\frac{2673}{14}\ln(3)\Biggl)\nu^2
+\biggl(- \frac{389727504721}{43545600}
 +\frac{74888}{243}\ln(2)\nonumber\\
&-& \frac{7128}{7}\ln(3)-\frac{30809603}{786432}\pi^4
-\frac{3934568}{8505}\gamma +\frac{9118627045}{5308416}\pi^2\Biggl)\nu\nonumber\\ 
e_{\rm 6PN}^{\ln}(\nu)&=& -\frac{1967284}{8505}\nu+ \frac{24464}{135}\nu^3+\frac{39754}{105}\nu^2 +\frac{11}{3}\nu^2 a_7^{\ln 1} \,.
\end{eqnarray}
\end{widetext}

Note the complexity of this expansion compared with the simplicity of the PN expansion of the EOB $A$ potential, Eqs.~(\ref{eq2})--(\ref{eqs6}). In particular, while $A(u;\nu)$ is linear in $\nu$ up to the 3PN level included, and is only
quadratic in $\nu$ at the 4PN level, $E_B (x) / \mu$ has a $O(\nu^2)$ contribution at 2PN, a $O(\nu^3)$ one at 3PN, and 
 a $O(\nu^4)$ one at 4PN. This simplicity of the $\nu$ dependence of the EOB radial potential  $A(u;\nu)$ (which, however,
 contains the {\it same} information as the function  $E_B (x) $) is due to the remarkable cancellations (recalled above)
 taking place  when computing  $A(u;\nu)$ from the PN Hamiltonian. There is, currently, no deep explanation for these
 cancellations. However, it is tempting to interpret them along the following lines.  The GSF  (i.e. $\nu$-dependent)
 part, say $ A_{\rm GSF}(u,\nu)$, of the $A$ potential  [$A(u;\nu)\equiv 1- 2u + A_{\rm GSF}(u,\nu)$] enters, at lowest
 order, and when considering the zero angular momentum case, the full Hamiltonian  through the contribution $ H_{\rm GSF} = \mu A_{\rm GSF}(u,\nu)$.  Let us consider a
 certain term in $A_{\rm GSF}(u,\nu)$ of the form $a_{n+1,p} u^{n+1} \nu^p$, i.e. a term at the $n$-PN level, which is
 proportional to $\nu^p$ (with an integer $p \ge 1$). If we now insert the expressions for $\mu$ , $\nu$, and $u$
 in terms of the two masses $m_1, m_2$, and of the (EOB) radial coordinate $r$, we find that the contribution
 to $ H_{\rm GSF} = \mu A_{\rm GSF}(u,\nu)$ corresponding to $a_{n+1,p} u^{n+1} \nu^p$ is of the form
 $(m_1 m_2)^{p+1} (m_1+m_2)^{n-2p}/r^{n+1}$. When $n$ is an integer, the latter contribution to the Hamiltonian
 will be a {\it polynomial} in the two masses (as one would expect to be the case, at least for an Hamiltonian
 computed in a usual gauge, such as the Arnowitt-Deser-Misner one)
 if and only if the power $p$ of $\nu$ satisfies the inequality $2 p \le n$.
 In turn, this inequality says that: (i) at 1PN ($n=1$), there cannot exist a $p \ge 1$ satisfying this condition;
 (ii) at 2PN ($n=2$), $p$ can only be equal to $1$;   (ii) at 3PN ($n=3$), $p$ can still only be equal to $1$; while
 (iv)   at 4PN ($n=4$), $p$ can only take the values $1$ or $2$.  All these consequences of the above requirement of polynomiality in the masses (i.e. the inequality $2 p \le n$) exactly coincide with what was found, by explicit computations, at 1, 2, 3 and 4PN
 (namely, absence of a 1PN contribution, linearity in $\nu$ at 2PN and 3PN, and quadratic nonlinearity in $\nu$
 at 4PN).  This interpretation   cannot be considered as a real ``explanation''
 of the cancellations found in $A(u,\nu)$, because the EOB Hamiltonian has not been proven to be directly obtainable
 by solving Einstein's equations by some kind of usual perturbation theory which keeps the polynomiality in the masses
 apparent at each stage.   However, its phenomenological success in recovering known facts suggests it has some truth. 
 [Note that, by contrast, the function $E_B(x;\nu)$, which is {\it not} a Hamiltonian, does not exhibit polynomiality properties
 in the two masses, even if we generously interpret the variable $x$ as being essentially of the form $ G (m_1+m_2)/c^2 r$.
 For instance, the presence of a term linear in $ \nu$ in the coefficient of the 1PN contribution $e_{\rm 1PN} (\nu)$
 corresponds to a term in  $E_B(G (m_1+m_2)/c^2 r;\nu)$ proportional to $\mu \nu  (G (m_1+m_2)/c^2 r)^2 $ and therefore
 proportional to $(m_1 m_2)^2/(m_1+m_2)$.]
Using our interpretation of the simplicity of    $A(u;\nu)$ as a heuristic basis for discussing higher PN orders, it suggests that: (v) at 5PN ($n=5$), $p$ can only take the values $1$ or $2$ (like at 4PN); while (vi) at 6PN ($n=6$), $p$ can only take the values $1$,  $2$ and $3$. [The $5.5$PN 
 contribution has to be treated as being offscale, because it is anyway not polynomial in $1/r$.] In terms of the
 parametrisation introduced above of the $\nu$ dependence beyond the 1GSF level,  the ``predictions'' (v) and (vi)
 mean that   $a_{6}^{c1}(\nu)$  would not depend on $\nu$, while  $a_{7}^{c1}(\nu)$,  $ a_{7}^{\ln 1}(\nu)$ would be at most linear in $\nu$.  In addition, the argument given in the next subsection suggests that
 $  a_{6.5}(\nu)$ does not depend on $\nu$, i.e. that $  a_{6.5}^{c1}(\nu)$ vanishes.
 
 If these suggested restricted $\nu$ dependences of the EOB-potential expansion coefficients $a_n(\nu)$ turn out to be 
 confirmed, many of the contributions
 to the binding-energy function  $ E_B(x;\nu) $  displayed above (those containing high powers of $\nu$) will be already
 correctly predicted by our (1GSF-based) results. For instance, the terms in $\nu^4$, $\nu^5$ and $\nu^6$
 in the $\nu$-exact PN expansion of $ e_{\rm 6PN}^c (\nu)$  might already be the ones explicitly appearing in the equations written above.
 A partial confirmation of this prediction concerns the terms in  $ E_B(x;\nu) $ that are related to the terms
 in the center-of-mass Hamiltonian 
which are of maximal order in $\nu$ at  some PN approximation.
 It has been recently emphasized in Ref. \cite{Foffa:2013gja}  that, at the $n$th PN order, the $\nu^n$  terms in the
 $\mu$-rescaled center-of-mass Hamiltonian carry at most one power of  $G$ (see, e.g.,  \cite{Damour:2014jta}),
 and suffice to determine the terms proportional to $\nu^n$ in $e_{ n {\rm PN}}(\nu) $.  Starting
 from the closed-form, first-post-Minkowskian (i.e. $O(G)$-accurate) Hamiltonian of  Ref.  \cite{Ledvinka:2008tk},  we explicitly computed (for $n=1,2,3,4,5,6$) the $\nu^n$  terms in the expansion coefficients $e_{ n {\rm PN}}(\nu) $ 
 of the function  $ E_B(x;\nu) $  and verified that they agree with the corresponding terms 
 explicitly appearing in  Eqs. (\ref{eB6PN}) above: e.g.  $+\frac{1}{512}\nu^5$ at 5PN, and  $+\frac{2717}{6718464}\nu^6$
 at 6PN. [The corresponding 5PN result  in  \cite{Foffa:2013gja}, namely  $+\frac{3121}{32}\nu^5$, appears to be incorrect.]
 This agreement confirms the usefulness of the EOB formalism which automatically encodes this information in the
 very simple ($\nu$-independent) 1PN-accurate EOB radial potential $A_{\rm 1PN}(u) = 1-2u$. We have more generally proven that this agreement holds to all PN orders. [This shows  that, at the $n$-PN level, the power of $\nu$ in 
 $\nu a_{n+1}(\nu)=\sum_p a_{n+1,p}  \nu^p$,  satisfies the inequality $p \leq n-1$, which is generally weaker than the
 inequality $ 2 p \leq n$ conjectured above.]
 Let us also note a generalization
 of the above property to the subleading order in the $\nu$ expansion:  we have verified that
 the terms proportional to $\nu^{(n-1)}$ in 
 $e_{ n {\rm PN}}(\nu) $ are entirely encoded (for  $n=1,2,3,4$) in the simple ($\nu$-linear) 2PN-accurate EOB 
 radial potential $A_{\rm 2PN}(u;\nu) = 1-2u + 2 \nu u^3$. This property will also hold at 5PN and 6PN if the currently unknown 5PN and 6PN coefficients have a limited-degree polynomial dependence on $\nu$ (e.g. $a_6^{c1}(\nu)$ would
 have to involve $\nu^2$ to modify the $\nu^4$ term in the 5PN energy contribution $e_{ 5 {\rm PN}}(\nu) $).

\subsection{Possible effect of the second-order tail integral in the inner metric and value of the $5.5$ PN coefficient $a_{6.5}(\nu)$}

The presence of a conservative contribution in the inner metric (and thereby in the redshift variable)  at the {\it odd}
PN order $1/c^{11}$ (namely the 5.5PN contributions $h_{6.5}$, $a_{6.5}$, or $e_{\rm 5.5PN}$ in the results above) is a priori surprising because conservative effects are traditionally associated with
even powers of $1/c$, while odd powers of $1/c$ are usually associated with time-odd radiation-reaction effects.
Let us show here, by a reasoning directly based on PN theory, why the 5.5PN order is the first order at which a 
conservative effect can involve an odd power
of $1/c$, and why the corresponding term must precisely be of the form we found in our RWZ calculations above.

The first  point  is that (as found in \cite{Blanchet:1987wq}), starting at the 4PN level, there are hereditary effects
 (given by an integral over the past behavior of the source) in the {\it inner metric} of the system that enter 
 radiation-reaction.
 Moreover, Ref.~\cite{Blanchet:1987wq} found that the  hereditary effects in the inner metric are time asymmetric
 without being time anti-symmetric. As a consequence they
 have both a dissipative aspect (hereditary 
modification of the lowest-order ($O(1/c^5)$), Burke-Thorne \cite{Burke,Thorne:1969rba},  2.5PN radiation reaction) and a conservative one (related to the 
4PN,  $O(1/c^8)$, logarithmic contribution to $A(u,\nu)$ \cite{Damour:2009sm}).
[The PN bookkeeping here is that the lowest-order tail-related
hereditary integral discussed in \cite{Blanchet:1987wq} has a prefactor $ G M/c^3$, which combines with the
leading-order (2.5 PN, i.e. $1/c^5$) radiation-reaction to go to the 4PN level $1/c^8$.]
In addition,  Ref.~\cite{Blanchet:1992br} found that the hereditary integral entering the inner metric obtained in  \cite{Blanchet:1987wq}  was related to a similar hereditary integral (``radiative tail'') entering the {\it wave-zone} metric by energy balance between the system and radiative losses 
at infinity (see Sec. IIID of \cite{Blanchet:1992br}). 
As  the latter energy-balance link between tail effects in gravitational radiation at infinity and hereditary integrals in the inner metric must clearly be generally satisfied, one expects that
higher-order tail effects in gravitational radiation will have corresponding hereditary counterparts in the inner metric,
which will modify radiation reaction by integrals having both dissipative aspects and conservative ones.
[Like the leading-order hereditary effects, the higher-order hereditary effects being time asymmetric without
being time antisymmetric can generate both radiation-reaction-like and conservative effects
(thereby violating the usual PN lore).]

The next-to-leading-order quadrupolar tail effect arises from a mass-mass-quadrupole interaction \cite{Blanchet:1997jj},
and will carry a prefactor   $ (G M/c^3)^2$.  In that case, energy balance indicates that
the second-order wave-zone tail in the {\it radiative} quadrupole moment (as seen at infinity) of the system \cite{Blanchet:1997jj} will similarly correspond to a 
 time-asymmetric  modification of the leading-order (2.5 PN) radiation reaction. The PN bookkeeping 
now tells us that the  $ (G M/c^3)^2$ prefactor combines with $O(1/c^5)$ to go to  $O(1/c^{11})$, i.e. the 5.5 PN level.
[Dimensional analysis shows that other tail interactions, e.g. of the
 mass-quadrupole-quadrupole type, start contributing only 
 at higher PN levels: 6.5PN and higher.]
 Let us now show
in detail how this time asymmetric (but not time anti-symmetric) modification of radiation reaction gives rise precisely to 
the 5.5PN {\it conservative}  contributions to $h_{kk}(u)$ and $A(u;\nu)$ we found above.

  Denoting the radiative quadrupole moment (evaluated at the
 retarded time $U$) as $M_{ij}^{\rm rad}(U)$, 
 and using a superscript $(n)$ to denote a $n$th time derivative, the result
 of  \cite{Blanchet:1997jj} (which was expressed in terms of $U_{ij} \equiv M_{ij}^{\rm rad \, (2)}$) can be written as
 \begin{eqnarray}
 \label{tail2}
 &&M_{ij}^{\rm rad}(U) = M_{ij}(U) \nonumber\\
&& + 2 \,  \frac{G \cal{ M}}{c^3}  \int_0^{+\infty} d \tau M_{ij}^ {(2)} (U-\tau) \ln\left(\frac{c \tau}{2 r'_0}\right) \nonumber \\
&&  +  2 \left(\frac{G \cal{ M}}{c^3} \right)^2 \int_0^{+\infty} d \tau M_{ij}^ {(3)} (U-\tau) \times \nonumber\\
&& \times \left[  \ln^2\left(\frac{c \tau}{2 r'_0}\right) + 
B  \ln\left(\frac{c \tau}{2 r'_0}\right) +C  \right] + \ldots \nonumber\\
&& 
+ {\rm semi-hereditary\, terms}
 \end{eqnarray}
Here,  $ \cal{ M}$ denotes the total mass-energy of the system,  $M_{ij}$ denotes the {\it algorithmic quadrupole moment}
of the Blanchet-Damour multipolar-post-Minkowskian formalism \cite{Blanchet:1985sp}, 
$r'_0 = r_0 \exp(-11/12)$ is a length scale differing from
 the basic length scale $r_0$ entering the latter formalism
by having absorbed the well-known
additional term $\frac{11}{12}$ entering the first-order radiative tail  \cite{Blanchet:1992br}, and the final ellipsis 
indicate some semi-hereditary (memory-type) terms that will not contribute to the near-zone effects discussed below
(which involve the 5th derivative of  $M_{ij}^{\rm rad}$). After the absorption of the latter $\frac{11}{12}$ in $r'_0$, the value of the
coefficient $B$ multiplying $ \ln(\frac{c \tau}{2 r'_0}) $ in the second-order hereditary term is
\be
B= - \frac{107}{105}\,.
\ee
The explicit expression given in Eq.~(\ref{Akup}) above for the crucial 5.5 PN coefficient $A_{11}^{{\rm up} \, (HG,\, l=2)}$ entering the
near-zone expansion of the ``up'' solution (and thereby the near-zone expansion of the Green's function,
and of the metric) shows that it  contains (notably in a term proportional to $X_1^2 X_2^{7/2} \propto \Omega^7 M^2$
which corresponds to what we discuss here) the first power of the logarithm of  $2\omega r \eta$, but not its second power.
This shows that the part of the second-order tail integral Eq.~(\ref{tail2})  involving a squared logarithm does not generate
a corresponding near-zone modification of radiation reaction. This is consistent with the finding of Blanchet  \cite{Blanchet:1997jj}  that, when computing the energy radiated at infinity by circular orbits, the  effect linked to the squared logarithm in the tail integral (\ref{tail2}) cancels with
a term in the square of the first-order tail integral. Let us note in passing that another way of understanding 
why the term $ \ln^2(\frac{c \tau}{2 r'_0}) $
in the tail integral does not contribute to the energy loss is via the scale dependence of the algorithmic moment  $M_{ij}$  \cite{Blanchet:1987wq,Blanchet:1997jj,Blanchet:2001aw,Goldberger:2009qd}: as the value of $M_{ij}$ depends on the arbitrary length scale $r'_0$,
but the physics does not depend on this choice, one can choose a scale $r'_0$  so as to simplify the result of the
tail integrals.  Choosing  $r'_0=r'_* :=c/(4 \Omega  \e^{\gamma})$ makes the first-order tail integral \cite{Blanchet:1993ec} 
$ \int_0^{+\infty} d \tau \exp (- 2 i \Omega \tau) \ln(\frac{c \tau}{2 r'_0})$ purely real, and makes the second-order
$\ln^2$ tail integral  $ \int_0^{+\infty} d \tau \exp (- 2 i \Omega \tau) \ln^2(\frac{c \tau}{2 r'_0})$ purely imaginary. One then sees, without doing any calculation
(taking into account the factors $( 2 i \Omega)^n$ implied by the $n$th derivatives acting on  $M_{ij}$) that this implies that
the energy loss must of the form $ \frac15 M_{ij}^ {(3)} [r'_*]  M_{ij}^ {(3)} [r'_*] \left( 1 + a \pi  \mathcal{ M}  \Omega + ( b  \pi^2 +c)  (\mathcal{ M} \Omega)^2 \right) $, with some rational coefficients $a, b, c$. The algorithmic
quadrupole moment $ M_{ij}[r'_*]$ used here corresponds to the special scale $r'_*$ defined above.
Using the 3PN logarithmic running of $ M_{ij}$ with the scale $r'_0$ discussed in  \cite{Blanchet:1987wq,Blanchet:1997jj,Blanchet:2001aw,Goldberger:2009qd}, namely
\be
\label{Mrunning}
 M_{ij}[r'_*] =   M_{ij}[r'_0]  + 2 B \ln \left(\frac{r'_*}{r'_0}\right) \left(\frac{G \mathcal{ M} }{c^3}\right)^2  M_{ij}^{(2)}[r'_0]  
 \ee
we can then obtain the energy loss for an arbitrary scale. We thereby find that it will  contain, at 3PN order,
only the first power of a logarithm (with a factor $\propto B  \mathcal{ M}^2$). This reasoning shows also why the coefficient of the logarithmic running is equal
(as we have just indicated) to $2 B$, where we recall that $B$ was defined as being the coefficient of the first power of the
logarithm in  the second-order tail in Eq.~(\ref{tail2}). Indeed,  when directly performing (as in \cite{Blanchet:1997jj}) the calculation of the energy loss
with an algorithmic moment corresponding to an arbitrary scale $r'_0$, the cancellation of the squared logarithms
leaves a term linear in $ \ln (\frac{r'_0}{r'_*}) = \ln (4 \Omega  \e^{\gamma} r'_0/c) $ with a coefficient equal to $16 B$.
This result is consistent with the reasoning above because the factor $16$ is the product of: the factor $2$ apparent in  Eq.~(\ref{Mrunning}), a factor $2^2$ from the
second derivative ($M_{ij}^{(2)} = - (2 \Omega)^2  M_{ij}$), and a factor $2$ from squaring $M_{ij}^{(3)}$.

The above reasoning (together with the results at first order in the tail  \cite{Blanchet:1987wq,Blanchet:1992br})
suggests that the {\it hereditary} effect of the second-order tail in the inner metric consists in modifying
the part of the inner metric connected to quadrupolar radiation reaction, i.e. 
\be
g_{00}^{\rm radreac}= - \frac{2G}{5c^7} x^i x^j  M_{ij}^{(5)}
\ee
by adding to $M_{ij}$ (taken at a scale corresponding to the size of the source) {\it twice} the contribution with
coefficient $B$  in  Eq.~(\ref{tail2}) (see Section III D of \cite{Blanchet:1992br} for a discussion of this factor 2).
  Note that the energy balance reasoning gives a handle only on hereditary terms,
and not on instantaneous terms, such as the term associated with the coefficient $C$ in Eq.~(\ref{tail2}).  Anyway, the
latter term is easily seen to be purely reactive (indeed it modifies  $M_{ij}^{(5)}$ into $  \sim C   \mathcal{ M}^2 M_{ij}^{(7)} $).
By contrast, the contribution involving the integral of  $ B \, \ln(\frac{c \tau}{2 r'_0})$, i.e.
\begin{eqnarray}
\label{tail2metric}
&& g_{00}^{{\rm tail}^2}= \\
&& - \frac{2G}{5c^7} x^i x^j   \left(  4 B \left(\frac{G \mathcal{ M}}{c^3} \right)^2 \int_0^{+\infty} d \tau M_{ij}^ {(8)} (U-\tau)   \ln\left(\frac{c \tau}{2 r'_0}\right) \right)\nonumber
\end{eqnarray}
contains conservative effects. Evaluating the tail integral along a circular orbit,  using (see \cite{Blanchet:1993ec})
\be
\int_0^{+\infty} dy \ln y  \,  \e^{- \sigma y} = -  \frac1\sigma (\ln \sigma + \gamma)
\ee
and computing the contribution of the metric  (\ref{tail2metric}) to the Hamiltonian, then leads to a contribution
(involving the logarithmic sum $S^{\log}_{7,2} $ )
to the EOB radial potential of a (comparable-mass) binary system equal to
\be
A^{{\rm tail}^2}(u, \nu) = - \frac{2^7}{5} B \pi  \nu u^{13/2} = + \frac{13696}{525} \pi \nu u^{13/2}\,.
\ee
In the limit $\nu \to 0$ this reproduces our result  Eq.~(\ref{eq6pnb}) above, but we get the new information that
$a_{6.5}(\nu)$ is actually independent of $\nu$. [This independence essentially follows from the facts that
$\mathcal{ M} = M + O(1/c^2)$ and that the Newtonian quadrupole moment is proportional to $\mu r_{12}^2$,
where the interbody distance $r_{12}$ is also related to $M$ and $\Omega$ by Kepler's law.]

Though the reasoning that led us to this result seems physically justified, let us emphasize that, at this stage,
it is only heuristic. One will need a  matching argument between the inner metric and the wave-zone one
(along the lines of Refs.~\cite{Blanchet:1987wq,Blanchet:2010zd}) to fully justify the final result. However, we find 
that the above reasoning clarifies the appearance of a conservative contribution at the 5.5 PN order, and strongly
suggests that our ($\nu$-linear) 1GSF result  Eq.~(\ref{eq6pnb}) remains valid for any value of $\nu$. 
[Ref. \cite{Blanchet:2013txa}, which appeared on the archives soon after our work, has confirmed the result of our 
analysis by  a detailed  multipolar-post-Minkowskian treatment of the gravitational field outside the source.]

\section{Conclusions}
\label{section:four}

Let us summarize our method and results:

We have studied several gauge-invariant functions characterizing the energetics of binary systems in
the limit $m_1 \ll m_2$.  At linear order in the (symmetric) mass ratio $\nu= m_1 m_2/(m_1+m_2)^2$,
these functions are linked by simple relations. On the one hand, Detweiler's redshift function  $ h_{kk}^{R}(u) := h_{\mu\nu}^{R} k^{\mu} k^{\nu}$ is related to its other avatars  through  Eqs. (\ref{huu}) and   (\ref{u1t}). On the other hand, it is related
to the $O(\nu)$ piece $a(u)$ in the main EOB radial potential $A(u;\nu)$,   Eq. (\ref{eq2a}),  through  Eq. (\ref{avshkk}).
Then, from $A(u;\nu)$ one can compute the function relating the binding energy of a binary to the orbital frequency
by  Eqs. (\ref{eq50}), (\ref{eq51}), (\ref{eq52}), (\ref{eq53}), (\ref{eq54}). [At linear order in $\nu$, the link between $h_{kk}$ and the binding energy was first derived in  \cite{Tiec:2011dp}; however, we prefer to use the EOB link between
 $A(u;\nu)$ and $E_B(x)$ because it is valid to all orders in $\nu$.]
 
 We used gravitational self force techniques to compute   $ h_{kk}^{R}(u)$ up to the 6PN accuracy, included. 
 The details of our method (which we already used in our previously reported 4PN-level computation) are spelled
 out in Sec. \ref{sec:two}. The final, 6PN accurate, result for  $ h_{kk}^{R}(u)$  is given in  Eq. (\ref{hkk6PN}). 
 The corresponding 6PN-accurate result for $a(u)$ is given in Eqs. (\ref{eq2a}) ,  (\ref{eq6pna}), (\ref{eq6pnb}), (\ref{eq6pnc}). Combining this information with the full $\nu$-dependent 4PN knowledge of    $A(u; \nu)$, and with a
 conjecture interpreting the remarkable cancellations of the $\nu$ dependence of   $A(u;\nu)$ up to 4PN
 (see the end of the previous Section), we obtained Eqs. (\ref{eB6PN}) for the 6PN-accurate energy
 function $E_B(u,\nu)$ (including part of the higher powers of $\nu$, notably the $\nu^n$ terms in $e_{n {\rm PN}}(\nu)$).
 
 An interesting result of our work is the finding of a 5.5PN contribution (i.e. of order $1/c^{11}$) 
  to the energetics of binary systems.
 As we discussed in the text, though this term, which is {\it conservative}, seems to conflict with the usual PN lore that conservative (time-even) effects arise at even powers of $1/c$, and that odd powers
of $1/c$ are associated with time-odd radiation-reaction effects, it is simply interpreted
in terms of the result of Blanchet and Damour \cite{Blanchet:1987wq}.  We have sketched the generalization of the  result of  \cite{Blanchet:1987wq}
to  second order in the tail factor \cite{Blanchet:1997jj} and found that it indeed explains the value we found (at the 1GSF approximation) for the
5.5 PN contribution to $A(u;\nu)$,  and suggests it is valid for any value of $\nu$. 
The important conceptual point here is that the hereditary
tail effects are time dissymetric without being time anti-symmetric. They can thereby generate terms that
are either radiation-reaction-like or conservative (thereby violating the usual PN lore).

Ref. \cite{Damour:2009sm} has shown that the 1GSF shift of the orbital frequency of the last stable (circular) orbit
was given (in terms of the dimensionless frequency parameter $x$) by $x_{\rm LSO}=\frac16 ( 1 + c_x \nu)$
(corresponding to $(m_1+m_2)\Omega_{\rm LSO}=6^{-3/2} ( 1 + c_{\Omega} \nu)$ with $c_{\Omega}= \frac32 c_x$),
where $c_x = \tilde a(1/6) + c_x^E$, with a universal part $c_x^E=\frac23 (1-\sqrt{ \frac89})$ and a part
depending on the derivatives of the EOB $a$ potential:
\be
 \tilde a(\frac16)=  a(\frac16) + \frac16 a'(\frac16) + \frac1{18} a''(\frac16)  \, .
 \ee
The numerical value of   $\tilde a(\frac16)$ was estimated by means of accurate GSF simulations in \cite{Akcay:2012ea},
and found to be
\be
 \tilde a^{\rm num}(\frac16) =0.795883004(15)
\ee
(corresponding to $c_{\Omega}^{\rm num}= 1.251015464(23)$).
By contrast, if we use our 6PN-accurate result (\ref{eq2c}) for $a(u)$, we find a value
\be
 \tilde a^{ 6PN}(\frac16) = 0.8495304315
 \ee
 (corresponding to   $c_{\Omega}^{6PN}=1.331486606$).

 As we see, the 6PN expansion of $\tilde a(u)$ yields an estimate of  $\tilde a(\frac16)$ which differs from  $\tilde a^{\rm num}(\frac16)$ by $6.7\%$.
 This is another example of the poor convergence of PN approximants in the (semi-) strong field regime (here, 
 $u= GM/c^2 r_0 = \frac16$). This poor convergence is linked with the presence of a (formal) singularity
 in the function $a(u)$ (and its redshift cousins, such as $h_{kk}$) at the light ring $u=\frac13$  \cite{Akcay:2012ea}.
 Indeed, this singularity implies that the PN expansion coefficients $a_n$ grow with $n$ like $\sim 3^n$. As a consequence
 the  $(n-1)$-PN contribution to $\tilde a(\frac16)$ coming from $a_n$  is expected to decrease, as $n$ increases, only  like $ (2 n^2 - n +1) a_n/6^n \sim 2 n^2 /2^n$ (see Eq. (4.35) in \cite{Damour:2009sm}). 
 For instance, the 6PN contribution linked to the non-logarithmic 
 coefficient $a_7 \simeq 738.28$ is equal to $0.24263$, which is far from being small, and actually represents $30 \%$
 of the final value of  $\tilde a(\frac16)$.

Let us finally comment on the comparison of our exact analytical higher-PN results to previous numerical
estimates. In terms of the coefficients of the PN expansion of $a(u)$, the numerical values of the new
coefficients we computed (including the previously reported 4PN one \cite{Bini:2013zaa}) are

\begin{eqnarray}
a_5&\approx & + 23.503389242603436240\nonumber\\
a_6&\approx & -134.07179509567553064\nonumber\\
a_{6.5}&\approx & +81.956672349649158466\nonumber\\
a_7&\approx & +738.2786376500962441\nonumber\\
a_7^{\ln}&= &-\frac{2522}{405}\approx -6.2271604938271604938\,. 
\end{eqnarray}
[We do not list here  the logarithmic coefficients $a_5^{\ln}$ and $ a_6^{\ln}$ that were analytically
known before, and that our work has confirmed.]
 
Previous accurate numerical investigations of Detweiler's redshift function led to corresponding
numerical estimates for these coefficients equal to \cite{Blanchet:2010zd,Barausse:2011dq}.
\begin{eqnarray}
a_5^{\rm num}&\approx &+ 23.50190(5)\nonumber\\
a_6^{\rm num}&\approx & -137.72(1)\nonumber\\
a_7^{\rm num}&\approx & +118(2)\nonumber\\
a_7^{\ln, {\rm num}}&= &-255.0(5)\,. 
\end{eqnarray}
The comparison between the two sets of values shows an increasing level of discrepancy as the
PN order increases. As already noted in  \cite{Bini:2013zaa}, even at the 4PN-level, the
analytical value of the coefficient $a_5$ differs by $30$ standard deviations from its previous
numerical estimate. The discrepancy becomes larger and larger for the other coefficients, to the
extent that they seem to be completely unrelated at the 6PN level. However, the reason for these
discrepancies is  clear. Ref.   \cite{Blanchet:2010zd} had not allowed for a 5.5PN contribution,
i.e. had effectively set $a_{6.5}$ to zero. As a consequence their fit had to absorb the missing 5.5PN
term by different values of the 6PN coefficients  $a_7^{\rm num}, 
a_7^{\ln, {\rm num}}$.  The missing ``signal'' associated with   $a_{6.5} u^{6.5}$ had also an effect on the
determination of the lower PN coefficients  $a_5^{\rm num}$, and $a_6^{\rm num}$ (with, as expected,
a stronger effect on the closest PN level). To check the idea that the discrepancies between our analytical estimates
and the previous analytical ones are entirely due to the non inclusion of a 5.5PN term in the expansion,
we have directly compared our 6PN-accurate analytical prediction for the quantity $h_{uu}^R$ (which
was the quantity fitted in \cite{Blanchet:2010zd}) to the best corresponding fitting polynomial obtained
in the latter reference (with coefficients denoted $a_j$ and $b_j$ there). We found that the difference
 $h_{uu}^{\rm analyt} - h_{uu}^{\rm numeri}$ stayed within the error level for $h_{uu}$ estimated in \cite{Blanchet:2010zd}.

  The results reported above were obtained during the writing of our short report \cite{Bini:2013zaa}.
 The preprint of  Shah et al. \cite{Shah:2013uya} (which appeared on the archives just before the submittal of our work)
have very recently provided new, ultra-high-accuracy numerical results on Detweiler's redshift variable $ u_{(1GSF)}^t$.
They give also  numerical evidence, and an analytical argument, for a 5.5PN contribution  to  $ u_{(1GSF)}^t$. To compare our results with theirs, one must re-express
our analytical 5PN, 5.5PN and 6PN results in terms of the variable  $ u_{(1GSF)}^t$. For this variable (which is related via Eq. (\ref{u1t})
to  $h_{kk}$)  our analytical results for the 6PN-accurate expansion read
\begin{widetext}
\begin{eqnarray}
&&u_{(1GSF)}^t = -u-2 u^2-5 u^3+\left(-\frac{121}{3}+\frac{41}{32}\pi^2\right)u^4\nonumber\\
&&+\left(\frac{677}{512}\pi^2-\frac{128}{5}\gamma-\frac{256}{5}\ln(2)-\frac{1157}{15}-\frac{64}{5}\ln u\right)u^5\nonumber\\
&&+\left(\frac{1606877}{3150}-\frac{60343}{768}\pi^2+\frac{1912}{105}\gamma+\frac{7544}{105}\ln 2+\frac{956}{105}\ln u-\frac{243}{7}\ln3\right) u^6\nonumber\\
&& -\frac{13696}{525}\pi u^{13/2}\nonumber\\
&& +\left(\frac{17083661}{4050}+\frac{102512}{567}\gamma+\frac{51256}{567}\ln u 
 +\frac{372784}{2835}\ln 2-\frac{1246056911}{1769472}\pi^2+\frac{1215}{7}\ln 3 +\frac{2800873}{262144}\pi^4\right) u^7\,.
\end{eqnarray}
\end{widetext}
In the notation of   \cite{Blanchet:2010zd}, i.e.
\be
u_{(1GSF)}^t= \sum_j (\alpha_j + \beta_j \ln u) \, u^{j+1}
\ee 
the higher-order PN coefficients, including the non-logarithmic  4PN one derived in \cite{Bini:2013zaa},
read (with the index $j$ now denoting the PN level)
\begin{widetext}
\begin{eqnarray}
\alpha_4&=& \frac{677}{512}\pi^2-\frac{128}{5}\gamma-\frac{256}{5}\ln(2)-\frac{1157}{15} \nonumber\\
&=& -114.34895136757260295\ldots \nonumber\\
\alpha_5&=& \frac{1606877}{3150}-\frac{60343}{768}\pi^2+\frac{1912}{105}\gamma+\frac{7544}{105}\ln 2-\frac{243}{7}\ln3 \nonumber\\ 
&= & -243.176814464674307587 \ldots \nonumber\\
\beta_5&=&  + \frac{956}{105} \nonumber\\
&= & 9.1047619047619047619\ldots \nonumber\\
\alpha_{5.5}&=& -\frac{13696}{525}\pi\nonumber\\
&= & -81.9566723496491584647\ldots \nonumber \\
\alpha_6&=& \frac{17083661}{4050}+\frac{102512}{567}\gamma+\frac{372784}{2835}\ln 2-\frac{1246056911}{1769472}\pi^2+\frac{1215}{7}\ln 3 +\frac{2800873}{262144}\pi^4\nonumber\\
&= & -1305.00138107870965574\ldots \nonumber\\
\beta_6&=& + \frac{51256}{567} \nonumber\\
&= & 90.39858906525573192239\ldots \,.
\end{eqnarray}
\end{widetext}
These numerical values  of our analytical results  agree  with the values
determined numerically (to a very high accuracy)  in  \cite{Shah:2013uya}. Our analytical result for  $\alpha_{5.5}$
agrees also with their independent result (as well as that of  \cite{Blanchet:2013txa}). We note also that our fully
analytical derivation of the 6PN logarithmic coefficient $\beta_6$ agrees with its numerical-analytical determination
in  \cite{Shah:2013uya}.\\

\appendix

\section{Relation $\nu$ vs $l$ at higher orders in $\epsilon$}

In Eq. (\ref{eq38}) we have cited (from \cite{Mano:1996vt,Mano:1996mf,Mano:1996gn})  the first term  in the expansion of $\nu$ in powers of $\epsilon^2$.
Let us exhibit here (for the case $s=-2$ of interest for us) explicit analytical expressions
for the next two terms, of order $\epsilon^4$, and
$\epsilon^6$:
\be
\nu = l +\nu_2(l) \epsilon^2+\nu_4(l) \epsilon^4+\nu_6(l) \epsilon^6+O(\epsilon^8)
\ee
We find
\begin{widetext}
\begin{eqnarray}
\nu_2(l)&=& -\frac12 \frac{(15 l^4+30 l^3+28 l^2+13 l+24)}{ l (3+2 l) (1+2 l) (-1+2 l) (l+1)}\\
\nu_4(l)&=& -\frac{1}{8}\frac{\bar \nu_4}{(5+2l)(-1+l)(1+2l)^3(2+l)(-3+2l)(3+2l)^3(-1+2l)^3(l+1)^3l^3}\nonumber\\
\nu_6(l)&=& -\frac{1}{16}\frac{\bar \nu_6}{ (-5+2 l) (1+2 l)^5 (7+2 l) (5+2 l)^2 (-1+l)^2 (2+l)^2 (-3+2 l)^2 (3+2 l)^5 (-1+2 l)^5 (l+1)^5 l^5} + c_6( \delta^l_2 - \delta^{-l-1}_{2}) \,.\nonumber
\end{eqnarray}
where
\begin{eqnarray}
&&\bar \nu_4=18480 l^{16}+147840 l^{15}+456120 l^{14}+605640 l^{13}+8295 l^{12}-1096830 l^{11}-1678310 l^{10}-1520455 l^9\nonumber\\
&&\quad -1355518 l^8-1397512 l^7-1217380 l^6-733273 l^5+675625 l^4+1855326 l^3+850608 l^2-102816 l-51840 \nonumber\\
&& \bar \nu_6 =-12975033600 l-2216596603241 l^{16}-2756712215600 l^{15}+69152667840 l^2+276057468576 l^3-595527565392 l^4\nonumber\\
&&-3036767042412 l^5-2535956468604 l^6+2930850757357 l^7-3984546031837 l^{13}-3214558957703 l^{12}+6222948337606 l^8\nonumber\\
&&+3813852229717 l^9-3680916388051 l^{14}-1750882207589 l^{11}+345064441861 l^{10}+938280719684 l^{19}-1215830743322 l^{18}\nonumber\\
&&+9306705408 l^{28}+104552448 l^{30}+1568286720 l^{29}-509720642816 l^{23}+527411908408 l^{22}-525016804928 l^{24}\nonumber\\
&&-187730326784 l^{25}+2143419358696 l^{21}-2130563709104 l^{17}+2569889990822 l^{20}-1170321152 l^{26}\nonumber\\
&& +24173140992 l^{27}-3919104000\,.
\end{eqnarray}
and where the coefficient $c_6$ of the last, additional term in $\nu_6(l)$
involving the Kronecker symbols $\delta^l_2$ and  $\delta^{-l-1}_{2}$ (which only contribute when $l=2$ or $l=-3$) is equal to
\be
c_6= \frac{14}{1605}
\ee
As pointed out to us by Ryuichi Fujita (private communication) the presence (say in the case $l=2$) 
of such an additional term is linked to the irregular behavior (as functions of $\epsilon$)
 of the coefficients $\alpha_n^{\nu}$ and $\gamma_n^{\nu}$ for the negative values $n=-l-1$
and $n = -l$ of the index $n$.
[Note that the term $O(\epsilon^6)$ corresponds to the 9 PN order, and did not enter the derivation
of the results reported here.]
The functions $\nu_{2k}(l)$ ($k=1,2,3$) are antisymmetric around $l=-\frac12$, i.e. they satisfy
\be 
\nu_{2k}(l)+\nu_{2k}(-l-1)=0\,.
\ee
In the cases $l=2,3,4$ used in the present work their values are listed below.
\be 
\begin{array}{|l||c|c|c|}
\hline
&&& \cr
l & \nu_2(l) & \nu_4(l) & \nu_6(l)\cr
&&& \cr
\hline
&&& \cr
2 &-\frac{107}{210} &-\frac{1695233}{9261000} & -\frac{76720109901233}{480698687700000}\cr
&&& \cr
\hline
&&& \cr
3 &-\frac{13}{42} &-\frac{10921}{271656} & -\frac{95353832269}{7709149047600}\cr
&&& \cr
\hline
&&& \cr
4 &-\frac{1571}{6930} &-\frac{68482418369}{4326563241000} & -\frac{6966001672062143707}{2701175970905111700000}\cr
&&& \cr
\hline
\end{array}
\ee
\end{widetext}

\noindent {\bf Acknowledgments.} T.D. thanks Francis Brown and Pierre Deligne for informative discussions
about the origin of the
transcendentality structure of the higher-PN coefficients $a_n(0)$, Steven Detweiler for clarifying email exchanges
about the regularization of $h_{kk}$, and Luc Blanchet for discussions about higher-order tail effects involving several multipoles.
We thank Ryuichi Fujita for informative email exchanges about the relation $\nu$ vs $l$ at higher orders in $\epsilon$,
and Stefano Foffa for attracting our attention towards the $\nu^n$ terms in $e_{n {\rm PN}}$.  
We are grateful to ICRANet for partial support. D.B. thanks IHES for hospitality during crucial stages of development of this project.


\begin{thebibliography}{99}

\bibitem{Bini:2013zaa} 
  D.~Bini and T.~Damour,
  ``Analytical determination of the two-body gravitational interaction potential at the 4th post-Newtonian approximation,''
  Phys.\ Rev.\ D {\bf 87}, 121501(R) (2013)
  arXiv:1305.4884 [gr-qc].

\bibitem{Buonanno:1998gg} A.~Buonanno and T.~Damour,
``Effective one-body approach to general relativistic two-body dynamics,''
Phys.\ Rev.\ D {\bf 59}, 084006 (1999)
[gr-qc/9811091].

\bibitem{Buonanno:2000ef} A.~Buonanno and T.~Damour,
``Transition from inspiral to plunge in binary black hole coalescences,''
Phys.\ Rev.\ D {\bf 62}, 064015 (2000)
[gr-qc/0001013].

\bibitem{Damour:2000we} T.~Damour, P.~Jaranowski and G.~Schaefer,
``On the determination of the last stable orbit for circular general relativistic binaries at the third postNewtonian approximation,''
Phys.\ Rev.\ D {\bf 62}, 084011 (2000)
[gr-qc/0005034].

\bibitem{Damour:2001tu} T.~Damour,
``Coalescence of two spinning black holes: an effective one-body approach,''
Phys.\ Rev.\ D {\bf 64}, 124013 (2001)
[gr-qc/0103018].

\bibitem{Jaranowski:2013lca} P.~Jaranowski and G.~Sch\"afer,
``Dimensional regularization of local singularities in the 4th post-Newtonian two-point-mass Hamiltonian,''
Phys. Rev. D {\bf 87}, 081503(R) (2013)
arXiv:1303.3225 [gr-qc].

\bibitem{Barausse:2011dq} E.~Barausse, A.~Buonanno and A.~Le Tiec,
``The complete non-spinning effective-one-body metric at linear order in the mass ratio,''
Phys.\ Rev.\ D {\bf 85}, 064010 (2012)
[arXiv:1111.5610 [gr-qc]].

\bibitem{Tiec:2011ab} A.~Le Tiec, L.~Blanchet and B.~F.~Whiting,
``The First Law of Binary Black Hole Mechanics in General Relativity and Post-Newtonian Theory,''
Phys.\ Rev.\ D {\bf 85} (2012) 064039
[arXiv:1111.5378 [gr-qc]].

\bibitem{Tiec:2011dp} A.~Le Tiec, E.~Barausse and A.~Buonanno,
``Gravitational Self-Force Correction to the Binding Energy of Compact Binary Systems,''
Phys.\ Rev.\ Lett.\  {\bf 108}, 131103 (2012)
[arXiv:1111.5609 [gr-qc]].

\bibitem{Detweiler:2008ft} S.~L.~Detweiler,
 ``A Consequence of the gravitational self-force for circular orbits of the Schwarzschild geometry,''
Phys.\ Rev.\ D {\bf 77}, 124026 (2008)
[arXiv:0804.3529 [gr-qc]].

\bibitem{Barack:1999wf} L.~Barack and A.~Ori,
``Mode sum regularization approach for the selfforce in black hole space-time,''
Phys.\ Rev.\ D {\bf 61}, 061502 (2000)
[gr-qc/9912010].

\bibitem{Mano:1996vt} S.~Mano, H.~Suzuki and E.~Takasugi,
``Analytic solutions of the Teukolsky equation and their low frequency expansions,''
Prog.\ Theor.\ Phys.\  {\bf 95}, 1079 (1996)
[gr-qc/9603020].

\bibitem{Mano:1996mf} S.~Mano, H.~Suzuki and E.~Takasugi,
``Analytic solutions of the Regge-Wheeler equation and the postMinkowskian expansion,''
Prog.\ Theor.\ Phys.\  {\bf 96}, 549 (1996)
[gr-qc/9605057].

\bibitem{Mano:1996gn} S.~Mano and E.~Takasugi,
``Analytic solutions of the Teukolsky equation and their properties,''
Prog.\ Theor.\ Phys.\  {\bf 97}, 213 (1997)
[gr-qc/9611014].

\bibitem{Sago:2002fe} N.~Sago, H.~Nakano and M.~Sasaki,
``Gauge problem in the gravitational selfforce. 1. Harmonic gauge approach in the Schwarzschild background,''
Phys.\ Rev.\ D {\bf 67}, 104017 (2003)
[gr-qc/0208060].

\bibitem{Nakano:2003he} H.~Nakano, N.~Sago and M.~Sasaki,
``Gauge problem in the gravitational selfforce. 2. First postNewtonian force under Regge-Wheeler gauge,''
Phys.\ Rev.\ D {\bf 68}, 124003 (2003)
[gr-qc/0308027].

\bibitem{Hikida:2004jw} W.~Hikida, S.~Jhingan, H.~Nakano, N.~Sago, M.~Sasaki and T.~Tanaka,
``A New analytical method for self-force regularization. 
II. Testing the efficiency for circular orbits,''
Prog.\ Theor.\ Phys.\  {\bf 113}, 283 (2005)
[gr-qc/0410115].

\bibitem{Hikida:2004hs} W.~Hikida, H.~Nakano and M.~Sasaki,
``Self-force regularization in the Schwarzschild spacetime,''
Class.\ Quant.\ Grav.\  {\bf 22}, S753 (2005)
[gr-qc/0411150].

\bibitem{Blanchet:2010zd} L.~Blanchet, S.~L.~Detweiler, A.~Le Tiec and B.~F.~Whiting,
``High-Order Post-Newtonian Fit of the Gravitational Self-Force for Circular Orbits in the Schwarzschild Geometry,''
Phys.\ Rev.\ D {\bf 81}, 084033 (2010)
[arXiv:1002.0726 [gr-qc]].

\bibitem{Shah:2013uya} 
  A.~G.~Shah, J.~L.~Friedman and B.~F.~Whiting,
  ``Finding high-order analytic post-Newtonian parameters from a high-precision numerical self-force calculation,''
  arXiv:1312.1952 [gr-qc].

\bibitem{Damourlogs} T.~Damour, 2010, unpublished; cited in  
  L.~Barack, T.~Damour and N.~Sago,
  ``Precession effect of the gravitational self-force in a Schwarzschild spacetime and the effective one-body formalism,''
  Phys.\ Rev.\ D {\bf 82}, 084036 (2010)
  [arXiv:1008.0935 [gr-qc]],
which quoted and used some combinations of the logarithmic contributions to $a(u)$ and $\bar d (u)$.
 
\bibitem{Damour:2009sm} T.~Damour,
``Gravitational Self Force in a Schwarzschild Background and the Effective One Body Formalism,''
Phys.\ Rev.\ D {\bf 81}, 024017 (2010)
[arXiv:0910.5533 [gr-qc]].

\bibitem{Blanchet:1987wq} L.~Blanchet and T.~Damour,
``Tail Transported Temporal Correlations In The Dynamics Of A Gravitating System,''
Phys.\ Rev.\ D {\bf 37}, 1410 (1988).

\bibitem{Blanchet:1992br} 
  L.~Blanchet and T.~Damour,
  ``Hereditary effects in gravitational radiation,''
  Phys.\ Rev.\ D {\bf 46}, 4304 (1992).

\bibitem{Blanchet:1997jj} 
  L.~Blanchet,
 ``Gravitational wave tails of tails,''
  Class.\ Quant.\ Grav.\  {\bf 15}, 113 (1998)
  [Erratum-ibid.\  {\bf 22}, 3381 (2005)]
  [gr-qc/9710038].
  
\bibitem{Blanchet:2013txa} 
  L.~Blanchet, G.~Faye and B.~F.~Whiting,
  ``Half-integral conservative post-Newtonian approximations in the redshift factor of black hole binaries,''
  arXiv:1312.2975 [gr-qc].

\bibitem{Blanchet:2009sd} L.~Blanchet, S.~L.~Detweiler, A.~Le Tiec and B.~F.~Whiting,
``Post-Newtonian and Numerical Calculations of the Gravitational Self-Force for Circular Orbits in the Schwarzschild Geometry,''
Phys.\ Rev.\ D {\bf 81}, 064004 (2010)
[arXiv:0910.0207 [gr-qc]].

\bibitem{Akcay:2012ea} S.~Akcay, L.~Barack, T.~Damour and N.~Sago,
``Gravitational self-force and the effective-one-body formalism between the innermost stable circular orbit and the light ring,''
Phys.\ Rev.\ D {\bf 86}, 104041 (2012)
[arXiv:1209.0964 [gr-qc]].

\bibitem{Detweiler:2002gi} S.~L.~Detweiler, E.~Messaritaki and B.~F.~Whiting,
``Selfforce of a scalar field for circular orbits about a Schwarzschild black hole,''
Phys.\ Rev.\ D {\bf 67}, 104016 (2003)
[gr-qc/0205079].

\bibitem{Sago:2008id} N.~Sago, L.~Barack and S.~L.~Detweiler,
``Two approaches for the gravitational self force in black hole spacetime: Comparison of numerical results,''
Phys.\ Rev.\ D {\bf 78}, 124024 (2008)
[arXiv:0810.2530 [gr-qc]].

\bibitem{Zerilli:1971wd} 
  F.~J.~Zerilli,
  ``Gravitational field of a particle falling in a schwarzschild geometry analyzed in tensor harmonics,''
  Phys.\ Rev.\ D {\bf 2}, 2141 (1970).

\bibitem{Barack:2005nr} 
  L.~Barack and C.~O.~Lousto,
  ``Perturbations of Schwarzschild black holes in the Lorenz gauge: Formulation and numerical implementation,''
  Phys.\ Rev.\ D {\bf 72}, 104026 (2005)
  [gr-qc/0510019].

\bibitem{Detweiler:2003ci} 
  S.~L.~Detweiler and E.~Poisson,
  ``Low multipole contributions to the gravitational selfforce,''
  Phys.\ Rev.\ D {\bf 69}, 084019 (2004)
  [gr-qc/0312010].

\bibitem{Chandrasekhar:1975} 
S. Chandrasekhar, Proc. R. Soc. London {\bf A 343}, 289 (1975).  

\bibitem{Blanchet:1984wm} 
  L.~Blanchet and T.~Damour,
  ``Multipolar radiation reaction in general relativity,''
  Phys.\ Lett.\ A {\bf 104}, 82 (1984).

\bibitem{Poisson:1994yf} 
  E.~Poisson and M.~Sasaki,
  ``Gravitational radiation from a particle in circular orbit around a black hole. 5: Black hole absorption and tail corrections,''
  Phys.\ Rev.\ D {\bf 51}, 5753 (1995)
  [gr-qc/9412027].

\bibitem{Sasaki:2003xr} 
  M.~Sasaki and H.~Tagoshi,
  ``Analytic black hole perturbation approach to gravitational radiation,''
  Living Rev.\ Rel.\  {\bf 6}, 6 (2003)
  [gr-qc/0306120].
  
\bibitem{Anderson:1982fk} 
  J.~L.~Anderson, L.~S.~Kegeles, R.~G.~Madonna and R.~E.~Kates,
  ``Divergent Integrals Of Postnewtonian Gravity: Nonanalytic Terms In The Near Zone Expansion Of A Gravitationally Radiating System Found By Matching,''
  Phys.\ Rev.\ D {\bf 25}, 2038 (1982).
  
  \bibitem{Damour:1988mr} T.~Damour and G.~Schaefer,
``Higher Order Relativistic Periastron Advances And Binary Pulsars,''
Nuovo Cim.\ B {\bf 101}, 127 (1988).

\bibitem{Jaranowski:1997ky}  P.~Jaranowski and G.~Schaefer,
``Third postNewtonian higher order ADM Hamilton dynamics for two-body point mass systems,''
Phys.\ Rev.\ D {\bf 57}, 7274 (1998)
[Erratum-ibid.\ D {\bf 63}, 029902 (2001)]
[gr-qc/9712075].

\bibitem{Damour:2001bu} T.~Damour, P.~Jaranowski and G.~Schaefer,
``Dimensional regularization of the gravitational interaction of point masses,''
Phys.\ Lett.\ B {\bf 513}, 147 (2001)
[gr-qc/0105038].

\bibitem{Jaranowski:2012eb} P.~Jaranowski and G.~Schafer,
``Towards the 4th post-Newtonian Hamiltonian for two-point-mass systems,''
Phys.\ Rev.\ D {\bf 86}, 061503 (2012)
[arXiv:1207.5448 [gr-qc]].

\bibitem{Damour:1995kt} 
  T.~Damour and G.~Esposito-Farese,
  ``Testing gravity to second postNewtonian order: A Field theory approach,''
  Phys.\ Rev.\ D {\bf 53}, 5541 (1996)
  [gr-qc/9506063].
  
\bibitem{Foffa:2013gja} 
  S.~Foffa,
  ``Gravitating binaries at 5PN in the post-Minkowskian approximation,''
  Phys.\ Rev.\ D {\bf 89}, 024019 (2014)
  [arXiv:1309.3956 [gr-qc]].

\bibitem{Damour:2014jta} 
  T.~Damour, P.~Jaranowski and G.~Sch\"afer,
  ``Non-local-in-time action for the fourth post-Newtonian conservative dynamics of two-body systems,''
  arXiv:1401.4548 [gr-qc].

\bibitem{Ledvinka:2008tk} 
  T.~Ledvinka, G.~Sch\"afer and J.~Bicak,
  ``Relativistic Closed-Form Hamiltonian for Many-Body Gravitating Systems in the Post-Minkowskian Approximation,''
  Phys.\ Rev.\ Lett.\  {\bf 100}, 251101 (2008)
  [arXiv:0807.0214 [gr-qc]].



\bibitem{Burke} 
  W.~L.~Burke,
  ``Gravitational Radiation Damping Of Slowly Moving Systems Calculated Using Matched Asymptotic Expansions,''
  J.\ Math.\ Phys.\  {\bf 12}, 401 (1971).
  
\bibitem{Thorne:1969rba} 
  K.~S.~Thorne,
  ``Nonradial Pulsation of General-Relativistic Stellar Models.IV. The Weakfield Limit,''
  Astrophys.\ J.\  {\bf 158}, 997 (1969).


\bibitem{Blanchet:1985sp} 
  L.~Blanchet and T.~Damour,
  ``Radiative gravitational fields in general relativity I. general structure of the field outside the source,''
  Phil.\ Trans.\ Roy.\ Soc.\ Lond.\ A {\bf 320}, 379 (1986).
  
\bibitem{Blanchet:2001aw} 
  L.~Blanchet, B.~R.~Iyer and B.~Joguet,
  ``Gravitational waves from inspiralling compact binaries: Energy flux to third postNewtonian order,''
  Phys.\ Rev.\ D {\bf 65}, 064005 (2002)
  [Erratum-ibid.\ D {\bf 71}, 129903 (2005)]
  [gr-qc/0105098].

\bibitem{Goldberger:2009qd} 
  W.~D.~Goldberger and A.~Ross,
  ``Gravitational radiative corrections from effective field theory,''
  Phys.\ Rev.\ D {\bf 81}, 124015 (2010)
  [arXiv:0912.4254 [gr-qc]].
  
\bibitem{Blanchet:1993ec} 
  L.~Blanchet and G.~Schaefer,
  ``Gravitational wave tails and binary star systems,''
  Class.\ Quant.\ Grav.\  {\bf 10}, 2699 (1993).
  
\end{thebibliography}
\end{document}